\documentclass[11pt,onecolumn]{article}
\usepackage{textcomp}
\DeclareMathAlphabet\mathbfcal{OMS}{cmsy}{b}{n}

\usepackage{hyperref}

\usepackage[linesnumbered,vlined,ruled]{algorithm2e}
\usepackage{amsmath,amsfonts,amssymb,amsthm}
\theoremstyle{plain}
\newtheorem{thm}{Theorem}

\newtheoremstyle{cited}%
  {3pt}
  {3pt}
{\itshape}
  {}
  {\bfseries}
  {.}
  {.5em}
  {\thmname{#1} \thmnumber{#2} \thmnote{\normalfont#3}}
\theoremstyle{cited}
\newtheorem{citedthm}[thm]{Theorem}
\newtheorem{citedlem}[thm]{Lemma}

\newtheorem{citedcor}[thm]{Corollary}

\usepackage{graphicx}
\usepackage{epstopdf}
\usepackage{epsf}
\usepackage{subfigure}
\usepackage{epsfig}

\usepackage{cite}

\usepackage{tabularx}
\usepackage{array}
\usepackage{booktabs}

\usepackage{comment}
\usepackage{tgpagella} 
\usepackage{mathpazo} 

\usepackage{algorithmic}
\usepackage{multirow}

\begin{document}

\title{Deriving RIP sensing matrices for sparsifying dictionaries}

\author{Jinn Ho and Wen-Liang Hwang}

\date{}
\maketitle

\begin{abstract}
Compressive sensing  involves the inversion of a mapping $SD \in \mathbb R^{m \times n}$, where $m < n$, $S$ is a sensing matrix, and $D$ is a sparisfying dictionary. The restricted isometry property is a powerful sufficient condition for the inversion that guarantees the recovery of high-dimensional sparse vectors from their low-dimensional embedding into a Euclidean space via convex optimization. However, determining whether $SD$ has the restricted isometry property for a given sparisfying dictionary is an NP-hard problem, hampering the application of compressive sensing. This paper provides a novel approach to resolving this problem. We demonstrate that it is possible to derive a sensing matrix for any sparsifying dictionary with a high probability of retaining the restricted isometry property.
In numerical experiments with sensing matrices for K-SVD, Parseval K-SVD, and wavelets, our recovery performance was comparable to that of benchmarks obtained using Gaussian and Bernoulli random sensing matrices for sparse vectors.
\end{abstract}
\pagenumbering{arabic}

\section{Introduction} \label{intro}

The Johnson-Lindenstrauss lemma~\cite{johnson1984extensions,dasgupta1999elementary} states that a set of points in a high-dimensional space can be embedded into a space of far lower dimensionality using an invertible Lipschitz function, such that distances between points are nearly preserved. Compressive sensing (CS) links this result within the framework of sparse representation. 
The problem of CS inversion involves the recovery of sparse vectors in a high-dimensional Euclidean space from their low-dimensional embedding via $SD$, where $S$ is a sensing matrix and $D$ a sparsifying dictionary. In accordance with the amount of a priori knowledge that is available, the issue of recovery has been formulated as a variety of problems
in which either both $S$ and $D$ are given (the CS recovery problem) or neither is given (simultaneously learning of sensing matrix and sparsifying dictionary~\cite{duarte2009learning}).

The restricted isometry property (RIP)\cite{candes2005decoding} of order $k$ assumes a restricted isometry constant $\delta_k \in [0, 1)$, such that for any $k$-sparse vector $x \in \mathbb R^n$,
\begin{align*} 
(1-\delta_k) \|x\|_2^2 \leq \|SDx\|_2^2 \leq (1+ \delta_k) \|x\|_2^2.
\end{align*}
If $\delta_k$ is small, then $SD$ is nearly linearly isometric, which means that it preserves the distance between any pair of $\frac{k}{2}$-sparse vectors.
Candes \cite{candes2008restricted} showed that k-sparse vectors can be correctly recovered via CS using convex optimization under the assumption that $SD$ has RIP with restricted isometry constant $\delta_{2k}$, where $k$ is the sparsity level when $\delta_{2k} < \sqrt 2 - 1$. The latter bound can be improved to $0.307$ \cite{cai2010new}. 
It has been demonstrated in~\cite{candes2006robust} that a stable inversion can be achieved for both $k$-sparse and compressible signals when $D$ is a basis (invertible square matrix) and that $SD$ has the RIP for $3k$ sparse vectors.
The main issue to utilize the  implications of the RIP is thus in constructing a sensing matrix $S$ for a given sparsifying dictionary $D$ such that $SD$ has the RIP.

There are sensing matrices that satisfy with high probability the RIP for a particular class of bases.
In \cite{candes2006robust,donoho2006compressed,baraniuk2006johnson}, the authors reported that a random matrix $S \in \mathbb R^{m \times n}$ with independently identically distributed (i.i.d.) entries (e.g.,  Gaussian and Bernoulli distributed) that satisfy the concentration inequality  is universal for any orthonormal matrix $D \in \mathbb R^{n \times n}$, in the sense that there is a high probability that $SD$ retains the RIP when $m \geq c k \log (\frac{n}{k})$, where $k$ is the sparsity level of vectors in $\mathbb R^n$ to be recovered.
Thus, a signal that is sparse with respect to some orthonormal basis can be coupled with random matrices to achieve the RIP. 

In practice however, one more commonly encounters signals that are sparse with respect to a redundant dictionary that may be far from orthonormal (e.g., Gabor-, curvelet-, wavelet-, or other data-driven learned dictionaries).
One solution to the above problem with respect to tight frames employs  the notion of D-RIP~\cite{candes2011compressed} 
\footnote{Let $\Sigma_k$ be the the union of all subspaces spanned by all subsets of $k$ columns of $D$.  The sensing matrix $A$ obeys the RIP adapted to $D$ with constant $\delta_k$ if
$
(1-\delta_k) \|x\|_2^2 \leq \|Ax\|_2^2 \leq (1 + \delta_k)\|x\|_2^2
$
holds for all $x \in \Sigma_k$.}. 
Assuming that the tight-frame coefficients $D^\top x$ are sparse, CS-recovery of $x$ from $y = Ax + \eta$, where $A \in \mathbb R^{m \times n}$ is a random Gaussian sensing matrix and the energy of the additive noise $\eta$ is bounded by $\|\eta\|_2<\epsilon$, can be achieved by solving
\begin{align} \label{DRIP}
\begin{cases}
\min_{x} \| D^\top x \|_1 \\
\|y - Ax \|_2 \leq \epsilon.
\end{cases}
\end{align}
It has been demonstrated in~\cite{candes2011compressed} that under the D-RIP framework, with high probability, it is possible to obtain stable inversion of $k$-sparse or compressible signals by solving \eqref{DRIP} in cases where the number of measurements $m$ is on the order of $k \log \frac{n}{k}$.

The nullspace property gives necessary and sufficient conditions on the reconstruction of sparse signals using the convex optimization \cite{cohen2009compressed}. The nullspace property is often difficult to check in practice.
Mutual incoherence between the sensing matrix $S$ and the dictionary $D$ is another frequently encountered condition that facilitates CS-recovery, albeit with greater restrictions on the sparsity level  of the vectors to be recovered than for the RIP~\cite{tropp2007signal,donoho2006compressed}.
In a given application, the prevalence of one sufficient condition over others is generally determined by a trade-off between the cost, memory capacity, and computational capability of the relevant devices.

This paper follows the line of the RIP in deriving a novel approach to solving the following problem:
Derive a sensing matrix $S$ for a given spasifying dictionary $D$, such that with high probability $SD$ retains the RIP.
It is widely believed that in seeking to satisfy the RIP, deriving random sensing matrices uses less number of 
sensing measurements than deriving deterministic sensing matrices \cite{rauhut2010compressive}.
Throughout the remainder of this paper, the symbol $\mathcal E$ is used to indicate the random-row-selection operation without repetition. 
We base our approach on the assumption that the dictionary $D \in \mathbb R^{l \times n}$, with $l \leq n$, takes the form of $GAH$, where $G \in \mathbb R^{l \times l}$ is invertible, $H \in \mathbb R^{n \times n}$ is orthonormal, and the probability distribution associated with the generation of $\mathcal EA \in \mathbb R^{m \times n}$ satisfies the concentration inequality of~\cite{dasgupta1999elementary,achlioptas2001database, krahmer2011new}, see Lemma~\ref{bara0}. 
The embedding function takes the usual form of $SD$, where $S$ refers to the sensing matrix. Clearly, if $\mathcal E G^{-1}$ is the sensing matrix of $D$, then $SD = \mathcal E AH$ has the RIP in the sense described in \cite{baraniuk2008simple}. 

Clearly, the success of the proposed approach depends on whether $D$ can be expressed as a $GAH$. Thus, the approach is referred to as the compressive sensing factorization approach to highlight the fact that a sparsifying dictionary is factorized into components that are used for compressing sensing purpose. We demonstrate that if the matrix $A$ (of the same rank of $D$) is also given, the answer to the problem is affirmative. In other words, given $D  \in \mathbb R^{l \times n}$ and $A \in \mathbb R^{l \times n}$, there exists an invertible matrix $G$ and orthonormal matrix $H$, such that $D= GAH$, provided that $D$ and $A$ have equal rank. If $l < n$, then the solutions $G$ and $H$ are not unique, due to the existence of a non-zero null space spanned by the rows of $A$. Despite their non-uniqueness, we demonstrate that both $G$ and $H$ can be determined analytically. In numerical experiments for the sensing matrices for K-SVD \cite{aharon2006k}, Parseval K-SVD\cite{hwang2019frame}, and CDF 9-7 wavelets \cite{cohen1992biorthogonal} we verify that the recovery performance of our approach is comparable to that of benchmarks achieved using Gaussian or Bernoulli random matrices as sensing matrices for sparse vectors.

The remainder of the paper is organized as follows. Section \ref{secframe} presents the background of CS. 
Section \ref{mainsec} presents the technical underpinnings of our approach and outlines the method by which we derive the sensing matrix for a given dictionary. Section \ref{experiment} presents experimental results on acquisition and representation using real-world images. Concluding remarks are presented in Section \ref{conclusion}.

\section{Background}\label{secframe}

The Johnson-Lindenstrauss lemma is concerned with embedding a set of $N$ points in $\mathbb R^n$ within a lower dimensional $\mathbb R^m$ (with $m$ is as small as possible), aiming to approximately preserve the distances between any two of the $N$ points.\footnote{Let $\epsilon \in (0, 1)$. For every set of $N$ points in $\mathbb R^n$ and $m > {\cal O}(\frac{\ln N}{\epsilon^2})$, there exists a Lipschitz mapping $f: \mathbb R^n \rightarrow \mathbb R^m$, such that for any two points $u$ and $v$ in the set, 
$
(1-\epsilon) \| u-v\|^2_2 \leq \| f(u) -  f(v) \|^2_2 \leq (1 + \epsilon) \| u - v\|^2_2.
$}
Various statements have been made based on this lemma, and a concise description of its evolution can be found in \cite{frankl1988johnson, indyk1998approximate}.
CS is linked to the Johnson-Lindenstrauss lemma in the framework of sparse representation, where the Lipschitz embedding is $SD$.

Sparse representations~\cite{chen2001atomic,donoho2006compressed,candes2006robust} seek to determine the sparsest coefficient vector (in terms of $\ell_0$-norm) from which signals can be synthesized using an over-complete dictionary. The $\ell_0$-norm objective is usually relaxed to the $\ell_1$-norm, in order to convert the non-convex into a convex optimization problem. The problem has gradually evolved from the recovery of sparse vectors based on observations, to deriving a sparsifying dictionary based on a set of observation \cite{aharon2006k,arora2014more,barak2015dictionary}.
In \cite{zhai2020complete}, learning an over-complete sparsifying dictionary after pre-conditioning empirical observations was reduced to learning an orthogonal transform, based on which the optimality of the solution and the complexity of the learning problem can theoretically be guaranteed. In \cite{schnass2018dictionary}, the parameters (i.e., the size of the dictionary and sparsity level) and the sparsifying dictionary are learned adaptively. There are numerous situations in which the adopted dictionary lacks the precision required for sparse representation. Pioneering work on the sensitivity of an imprecise dictionary to CS recovery was performed in \cite{herman2010mixed}. The sensitivity of the sparse recovery against dictionary perturbation in terms of the restricted isometry constant under the CS paradigm is outlined in \cite{ho178ambiguity}.

When using CS for signal acquisition, signals that have sparse coefficients with respect to $D$ can be mapped into a much lower-dimensional space via a sensing matrix $S$.
Despite the results in \cite{baraniuk2008simple} - indicating that orthonormal, Gaussian and Bernoulli sensing matrices retain the RIP -
deriving the composition of a dictionary with a sensing matrix that retains the RIP is an NP-hard problem and the main issue hindering the imposition of the RIP as a constraint in CS problems.
Nonetheless, researchers have devised ingenious dictionary and sensing matrix pairs as well as algorithms to facilitate CS recovery. 
It has been demonstrated that if the sparsity level $k$ of a vector $x$ is smaller than a constant multiple of the reciprocal of the coherence of $SD$, then it is possible to recover $x$ from $y = SD x$ \cite{donoho2003optimally}. Moreover, numerical experiments have demonstrated that CS works well even when the sparsity level exceeds the theoretically guaranteed bound. For example, Elad \cite{elad2007optimized} demonstrated that the optimal choice of sensing matrix can improve incoherence with a pre-defined sparsifying dictionary, thereby substantially improving CS reconstruction performance. A number of algorithms have been developed with the aim of recovering both the sensing matrix and the sparsifying dictionary from observations. For example, the method 
in \cite{duarte2009learning} can simultaneously learn the sensing matrix and sparsifying dictionary from given samples $\{(y_i, z_i) \}_{i=1}^N$, for which $y_i = D x_i$ and $z_i = S y_i + \eta$, where $\eta$ is additive noise contained in the sensing system.

From the perspectives of computation and implementation, it is desirable to use structured sensing matrices~\cite{devore2007deterministic}.
In \cite{rauhut2010compressive}, random-row-sampling $\mathcal E$ is used as the sensing matrix for any bounded orthonormal system.
A well-known example of this approach is the random partial Fourier matrix, in which rows of a Fourier matrix are randomly sampled without repetition. 
Random partial circular and Toeplitz matrices, obtained by applying $\mathcal E$ as a sensing matrix to circular and Toeplitz matrices, are used in wireless communication and radar applications~\cite{bajwa2007toeplitz}.
Romberg~\cite{romberg2009compressive} developed an approach in which $\mathcal E$ is used as the sensing matrix for a random square matrix of the form $D = (F^* \Sigma F)\psi \in \mathbb R^{n \times n}$, where $\psi$ is an orthonormal system, the entries of the diagonal matrix $\Sigma$ are randomly selected numbers within specified ranges, and $F$ is the Fourier matrix. 
The RIPless approach~\cite{candes2010probabilistic} employs a sensing mechanism capable of approximately recovering sparse signals from noisy measurements, as long as the random sensing measurements are selected independently from a probability distribution with specific properties.

The fact that Gaussian and Bernoulli sensing matrices do not provide rapid matrix multiplications or economical storage precludes their adoption as sensing matrices for large scale orthonormal systems. To the best of our knowledge, matrix factorization was first used for a sensing matrix in the structurally random method (SRM) \cite{do2011fast}, where a matrix product involving $\mathcal E$ is used as sensing matrix. 
The SRM makes it possible to construct a sensing matrix that can be calculated rapidly, thereby eliminating  the need to store the sensing matrix explicitly. This approach is applicable to any sensing matrix of the form $\mathcal EWR$, where $W$ is an orthonormal matrix and $R$ is either a random permutation or a diagonal matrix. The low mutual coherence of SRM sensing matrices with any orthonormal matrix makes them suitable sensing matrices for large scale orthonormal systems.

The conventional approach to solving an inverse problem is via algorithms that iteratively approximate a solution to within a desired error bound, in conjunction with a theoretical analysis to determine the rate of convergence. This approach can hamper applications that require rapid decisions, or expensive computations for each iteration. Deep neural networks (DNNs) have been used to solve inverse problems \cite{mousavi2017learning,monga2021algorithm} as well as the CS recovery problem, wherein a sparse vector, or a signal that allows for sparse transform coefficient, is derived via forward inference from a low-dimensional input. Numerical experiments have demonstrated that the DNN approach is well-suited to the CS recovery problem \cite{zhang2018ista,bora2017compressed,hwang2021learning}, allowing the recovery of a sparse vector using far fewer measurements than the conventional approach. Unfortunately, the DNN approach, so far, does not provide theoretical justification for the solution.

\section{Sensing matrix construction for prescribed dictionaries} \label{mainsec}
In this section we provide the details of our proposed approach. Recall that $\mathcal E$ denotes the random-row-selection operation without repetition.  In the following, let $A \in  \mathbb R^{l \times n}$ be a random matrix, and suppose $\mathcal EA \in \mathbb R^{m \times n}$ satisfies the concentration inequality of the following result of \cite{baraniuk2008simple}, which guarantees that with high probability $\mathcal EA$ has the RIP.

\begin{citedlem}[(Concentration inequality: \cite{baraniuk2008simple}, Theorem 5.2)] \label{bara0}
Suppose $\mathcal E \in \mathbb R^{m \times l}$ and $A \in \mathbb R^{l \times n}$. For any $x$, if the probability distribution generating $\mathcal EA \in \mathbb{R}^{m\times n}$ satisfies
$\mathbb E\|\mathcal EA x \|_2^2 = \| x\|_2^2$ and for some  $\epsilon \in (0, 1)$ the concentration inequality
$\mathbb{P}(|\|\mathcal EAx\|_2^2 - \|x\|_2^2| \geq \epsilon \|x\|_2^2) \leq 2 e^{- l c(\epsilon)}$ holds for some $c(\epsilon)>0$ depending only on $\epsilon$, where the probability is taken over all $A \in \mathbb R^{l \times n}$,
then there exist $c_1, c_2 > 0$ (depending only on $\delta_k$) such that the RIP, with the restricted isometry constant  $\delta_k$ and $k \leq \frac{c_1m}{ \log(n/k)}$, holds for $\mathcal EA$  with probability at least $1-2 e^{-c_2 m}$. 
\end{citedlem}
The random matrices whose entries are independent realizations of Gaussian random variables or Bernoulli random variables satisfy the above lemma \cite{baraniuk2008simple}.
It is clear that, if $\mathcal EA$ satisfies the concentration inequality, then so does $\mathcal EAH$ for any orthonormal $H$. The following corollary is therefore immediate.

\begin{citedcor} \label{bara2}
Suppose that $\mathcal E A$ satisfies the assumptions of Lemma \ref{bara0} and that the sparsifying dictionary $D \in \mathbb R^{l \times n}$ can be decomposed as $GAH$ for some invertible $G$ and orthonormal $H$. Then for the sensing matrix $S:=\mathcal E G^{-1} \in \mathbb R^{m \times l}$, the embedding $SD = \mathcal EAH$ has the RIP in the sense of Lemma~\ref{bara0}. 
\end{citedcor}

By the following lemma, the existence of a factorization $D = GAH$ with invertible $G$ and orthonormal $H$, as assumed in Corollary~\ref{bara2}, is equivalent to $D$ and $A$ having equal rank. 

\begin{citedlem}\label{PROPOSITION_1}
Let $A,D \in \mathbb{R}^{l\times n}$ with $l \leq n$. Then,
the following statements are equivalent: \footnote{(o) $\{WD \mid W\in \mathbb{R}^{l\times l} \text{ is invertible}\}
\cap\{AH \mid H\in \mathbb{R}^{n\times n} \text{ is orthonormal} \} \ne \emptyset$.}
\begin{itemize}
\item[(i)] There exists an invertible
$W\in \mathbb{R}^{l\times l}$ and  orthonormal $H\in \mathbb{R}^{n\times n}$ such that
$WD = AH$.
\item[(ii)]
There exists an invertible $W\in \mathbb{R}^{l\times l}$ such that
$AA^{\top} = 
WDD^{\top} W^{\top}$.
\item[(iii)]
$A$ and $D$ have equal rank.
\end{itemize}
\end{citedlem}
\proof
The implications (i)$\Rightarrow$(ii)$\Rightarrow$(iii) are straightforward.
To show (iii)$\Rightarrow$(ii), assume $A$ and $D$ have equal rank $k$. Since row-rank equals column-rank, also the symmetric matrices $AA^{\top}$ and $DD^{\top}$ have equal rank $k$. 
Thus the desired $W$ can be derived from the spectral decompositions
$Q_A  \Sigma_A Q_A^{\top}$ of $AA^\top$ and
$Q_D  \Sigma_D Q_D^{\top}$ of $DD^\top$,
where $\Sigma_A$ and $\Sigma_D$ are diagonal with $k$ nonzero entries, assumed to be in the same positions.
Since  $\text{ rank} (AA^{\top}) = \text{ rank} (DD^{\top}) $, 
there exists an invertible diagonal matrix $\Sigma_S$ such that
\begin{equation}\label{Prop_Sigma_S}
\Sigma_A  = \Sigma_S \Sigma_D \Sigma_S.
\end{equation} 
Then
$
W := Q_A  \Sigma_S Q_D^{\top}
$ 
is invertible, and 
\begin{align}\label{Prop_WDDW}
WDD^{\top} W^{\top}&=
W
Q_D  \Sigma_D Q_D^{\top} 
W^\top=  Q_A \Sigma_S \Sigma_D \Sigma_S Q_A^{\top} =   AA^{\top}. 
\end{align}

To show (ii)$\Rightarrow$(i), by the equivalence of (ii) and (iii) we can assume $A$ and $D$ have equal rank $k\le l$, and considering the spectral decompositions and $W$ as above.  We let $A^+$ to denote the Moore–Penrose pseudo-inverse of matrix $A$.
Let $N_A , N_D \in \mathbb{R}^{n\times (n-k)}$ consist of pairwise orthonormal columns spanning the null spaces of $A$, resp.\ $D$, i.e.,
$A N_A = D N_D = 0_{l \times (n-k)}$ and
$N_A^{\top}N_A =N_D^{\top}N_D =I_{n-k}$.
Let $H:=A^+WD+N_AN_D^\top$. Then
\[
AH = AA^+WD+AN_AN_D^\top =  AA^+WD = WD
\]
since $AA^+$ is the identity on the range of $A$, of which the range of $W$ is a subspace, by definition. Moreover, by \eqref{Prop_WDDW},
\[
HH^T = A^+WDD^\top W^\top(A^+)^\top + N_AN_A^\top = 
A^+ ( A A^\top (A^+)^\top) +  N_AN_A^\top = 
A^+ A + N_A N_A^\top = I_n. \qedhere
\]

\medskip
Note that the factorizations in Lemma~\ref{PROPOSITION_1} are not unique, since the spectral decompositions involved in their construction depend on the order in which the spectra are considered. Letting $G:=W^{-1}$, Lemma~\ref{PROPOSITION_1} implies the following result.

\begin{citedthm}\label{mainthm}
Suppose $A,D \in \mathbb{R}^{l\times n}$ ($l\leq n$) have equal rank $k$. Then $D= GAH$ for some invertible $G$ and orthonormal $H$ (and this decomposition is not unique). 
\end{citedthm}

\medskip
A sensing matrix $S$ for a given sparsifying dictionary $D$ can be constructed from any $A$ of equal rank that satisfies the concentration inequality, and any factorization $D=GAH$ as in Theorem~\ref{mainthm}. Indeed, letting $S := DG^{-1}$,  Corollary~\ref{bara2} then implies that the operator $SD = DAH$ has the RIP.

\medskip
Tight frames are dictionaries $D$ with the property that $DD^\top$ is a multiple of the identity. They are ubiquitous in signal and image processing~\cite{bao2013fast}, since then the condition number of $DD^\top$ is equal to $1$ and the canonical dual (used to derive decomposition coefficients of a signal) is given by $D^\top$ (up to a constant rescaling). If $D$ is a tight frame, we may assume $DD^\top=I_l$ (by rescaling the columns of $D$ if necessary). If $A$ and $D$ have full rank, a factorization as in Theorem~\ref{mainthm} can then be constructed as follows.\footnote{If $A,D$ have equal --  but not full --  rank, the following result holds replacing the inverses by the pseudo-inverses.}

\begin{citedthm} \label{tightframe}
Suppose $A,D \in \mathbb{R}^{l\times n}$ have full rank $l\leq n$, and $D$ is a tight frame. Then $D=GAH$ with invertible, respectively orthonormal, factors defined by
\begin{align*}
G := O(AA^\top)^{-1/2} 
\quad\text{ and }\quad
H := A^{\top}G^{\top} D +  N_A N_D^\top,
\end{align*}
where $O \in \mathbb{R}^{ l \times l}$ is an orthonormal matrix and the columns of $N_A , N_D \in \mathbb{R}^{n\times (n-l)}$ are any orthonormal bases of the null spaces of  $A$, resp.\ $D$.  
\end{citedthm}
\proof
Note first that
$GAA^{\top}G^{\top} =  I_l$ and thus $D=GAH$.
Further, $A N_A = D N_D = 0_{l \times (n-l)}$ and
$N_A^{\top}N_A =N_D^{\top}N_D =I_{n-l}$. Thus the
tight-frame property of $D$ implies that
\begin{align*}
HH^{\top} &= A^{\top}G^{\top} D D^{\top} G A 
+N_AN_A^{\top} = A^{\top}G^{\top} G A +N_AN_A^{\top}
=  A^{\top} (AA^{\top})^{-1} A +N_AN_A^{\top}
= I_n. \qedhere
\end{align*}

\medskip
We conclude this section by detailing one particular factorization according to Theorem \ref{mainthm}; based only on orthonormal bases for the ranges of $A$ and $D$: Given $A,D \in \mathbb{R}^{l\times n}$ of rank
$k\le l$, choose $U,V \in \mathbb{R}^{n\times k}$, each with pairwise orthonormal columns, such that
\begin{align}\label{GramSchmidt}
AUU^{\top} = A
\quad\text{ and }\quad
DVV^{\top} = D.
\end{align}
(Such $U$ and $V$ may be derived via Gram-Schmidt through constructing orthonormal column vectors spanning the ranges of $A$ and $D$.) 
Next consider 
$U_{\perp},V_{\perp} \in \mathbb{R}^{n\times (n-k)}$ such that the extended matrices $[U \,\, U_{\perp}],[V \,\, V_{\perp}]\in\mathbb{R}^{n\times n}$ are orthonormal,
and define
\begin{align}\label{Hdetermine}
H := UV^{\top} + U_{\perp}V_{\perp}^{\top}.
\end{align}
Then, clearly, $H$ is orthonormal and 
$AH = AUV^{\top}$.
The latter implies, by \eqref{GramSchmidt},
\begin{align*}
D - GAH = (DV-GAU)V^{\top}
\end{align*}
for any $G \in \mathbb{R}^{l\times l}$. 
Since $V^T$ has linearly independent rows, it follows that,
\begin{align}\label{iff}
D - GAH = 0 \quad\text{ if and only if }\quad DV-GAU = 0.
\end{align}
We extend $DV, AU \in  \mathbb{R}^{l\times k}$ (where $k\le l$) to invertible square matrices $\widehat{DV},\widehat{AU}$, appending columns.
Then
\begin{align} \label{Gdetermine}
G := \widehat{DV}(\widehat{AU})^{-1} \in  \mathbb{R}^{l\times l}
\end{align} 
is invertible with
$\widehat{DV} = G \widehat{AU}$, and thus
$DV=GAU$. The latter implies $D=GAH$ by \eqref{iff}.
Note that, again, the definition of $H$ via \eqref{Hdetermine} is not unique, but depends on the Gram-Schmidt procedure to derive $U,V$. Moreover, if $k < l$ the definition of $G$ via \eqref{Gdetermine} is not unique;  even for fixed $U,V$.

\section{Experimental results} \label{experiment}

\begin{figure}[h!]
{\centerline{\hspace{0.5cm}
\epsfig{figure=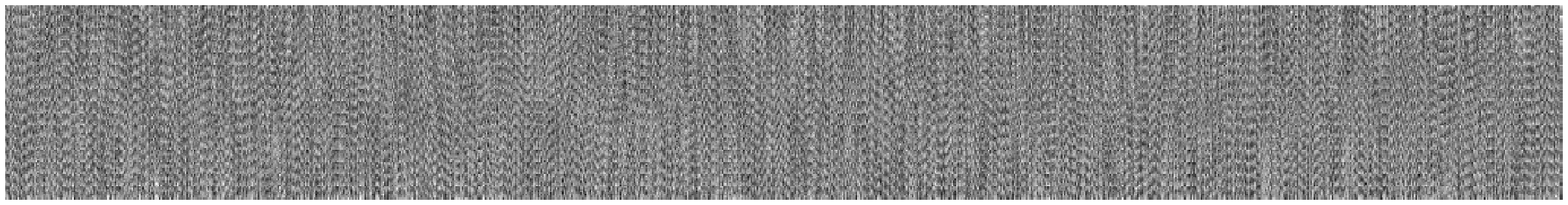,width=18cm}}}
\centerline{\hspace{0.8cm}(a) $D_{KSVD} \in \mathbb R^{128 \times 1024}$}
{\centerline{\hspace{0.5cm}
\epsfig{figure=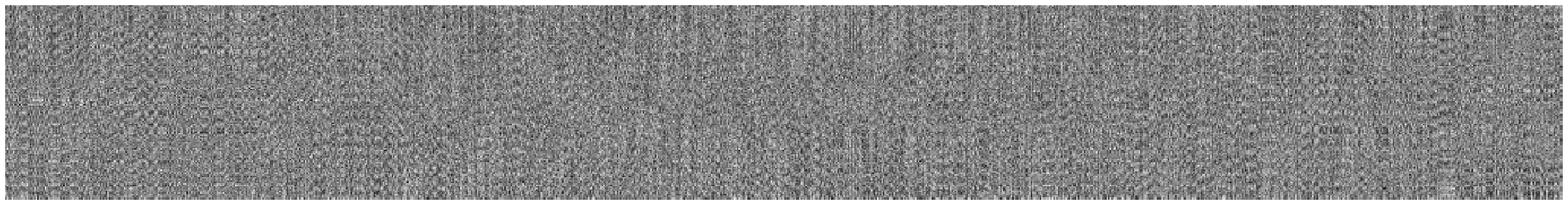,width=18cm}}}
\centerline{\hspace{0.8cm}(b) $D_{PKSVD}\in \mathbb R^{128 \times 1024}$ }
{\centerline{\hspace{0.5cm}
\epsfig{figure=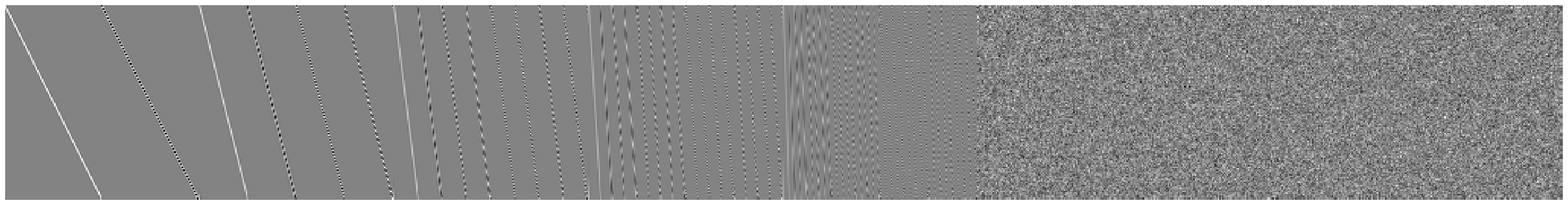,width=18cm}}}
\centerline{\hspace{0.8cm}(c) $D_{wavelet} \in \mathbb R^{128 \times 1024}$ }
\caption{\label{fig_D}
Visualization of sparsifying dictionaries with integer values ranging from $0$ to $255$.
}
\end{figure}

Experiments were conducted to compare CS recovery performance using the proposed compressive sensing factorization approach with benchmarks derived using Gaussian or Bernoulli random matrices as sensing matrices for sparse vectors. The entries of $A \in \mathbb R^{l \times n}$ in our experiments are i.i.d. Bernoulli and Gaussian random numbers where
the Gaussian random variables are $\mathcal N(0, n^{-1})$ and the Bernoulli random variables are $\pm \frac{1}{\sqrt{n}}$ with equal probability. In all experiments,
the sparse vectors were derived using the compressive sampling matched pursuit (CoSaMP) algorithm \cite{needell2009cosamp}. The formulation of CoSaMP for our approach, with sparsifying dictionary $D = GAH$, is
\begin{align}
\begin{cases}
\displaystyle \min_{x_i}  \| z_i - \mathcal E_i G^{-1} D x_i \|^2_2 \\
\| x_i  \|_0 \leq k, \label{cs_p1}
\end{cases}
\end{align}
while for the benchmark methods used for comparison it is
\begin{align}
\begin{cases}
\displaystyle \min_{x_i}  \| z_i - \mathcal E_i A x_i \|^2_2 \\
\| x_i  \|_0 \leq k  \label{cs_p2}.
\end{cases}
\end{align}
Note that $z_i$ in (\ref{cs_p1}) and (\ref{cs_p2}) are obtained from $\mathcal E_i A x_i$ and $\mathcal E_i$ in (\ref{cs_p1}) and (\ref{cs_p2}) are the same. Following the fact that if $D=GAH$ is plugged into (\ref{cs_p1}), the only difference is $AH$ in (\ref{cs_p1}) and $A$ in (\ref{cs_p2}) and $AH$ and $A$ have the same distribution, we can derive that applying sensing matrix $\mathcal E_i G^{-1}$ to dictionary $D$ for sparse recovery (\ref{cs_p1}) yields a performance similar to (in terms of probability) the sparse recovery using (\ref{cs_p2}). To verify this, we performed experiments on three sparsifying dictionaries in $\mathbb R^{128 \times 1024}$: $D_{wavelet}$, $D_{KSVD}$, and $D_{PKSVD}$ (a variant of K-SVD optimized to be a Parseval tight frame). The latter two dictionaries were respectively derived via the K-SVD and PK-SVD algorithms, using the same set of training vectors obtained from the $256 \times 256$ gray-scale test image ``Boat"\footnote{Parameters of the K-SVD process were set at $K=1024$,
$L=64$,
$numIteration=50$,
$InitializationMethod='GivenMatrix'$, and   
$errorGoal=10^{-8}$. 
Parameters for PK-SVD were set at
$\Phi_0=\Psi_0=D_{KSVD}$,
$L=64$,
$maxIter = 50$,
$\rho = [0.1, 10^{8}, 10^{8}]$, and
$t=10^{-10}$.
}. 
The image was divided into overlapping patches of $16 \times 16$ pixels with stride $2$ (in each dimension) with overlaps of $4$ pixels (in each dimension), which resulted in $3,721$ training patches. Since the stride is $2$ in each dimension, each patch forms a vector of size $128$. The mean of each patch was normalized to zero, whereupon the resulting patches were transformed into vectors (via the vec operation) to form the set of training vectors. The dictionary
$D_{wavelet}$ is derived from the CDF 9-7 wavelet. 
The first $640$ (i.e., each level has $128$ columns and there are $5$ levels) columns in $D_{wavelet}$ were obtained from the first $5$-level of the wavelet, whereas the remainder were generated randomly. The column norms of the sparsifying dictionaries were normalized. Figures~\ref{fig_D} and \ref{fig_G} present the sparsifying dictionaries and corresponding matrices $G$ for various $A$.

\begin{figure}[h!]
{\centerline{\hspace{0cm}
\epsfig{figure=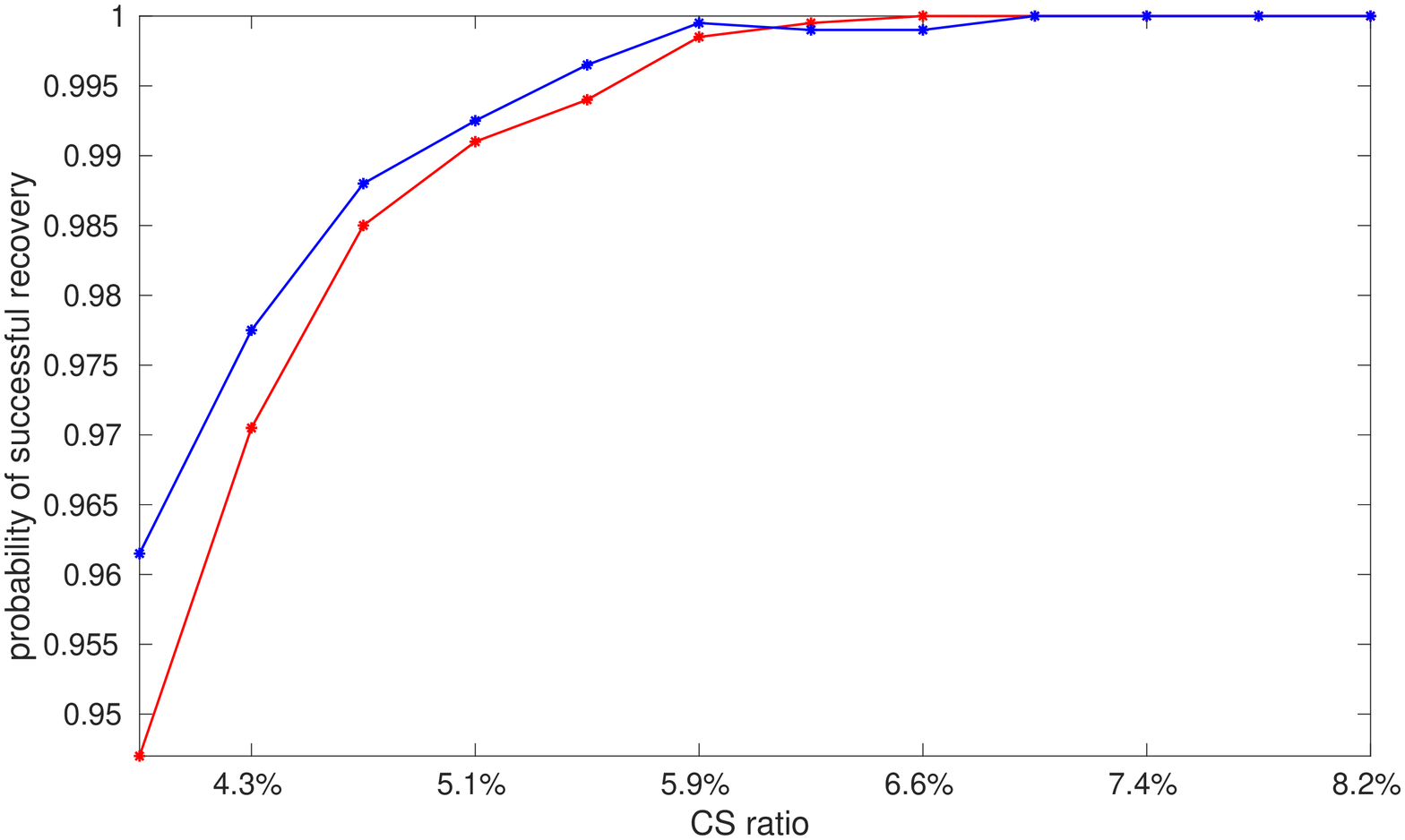,width=8cm}\hspace{0.4cm}
\epsfig{figure=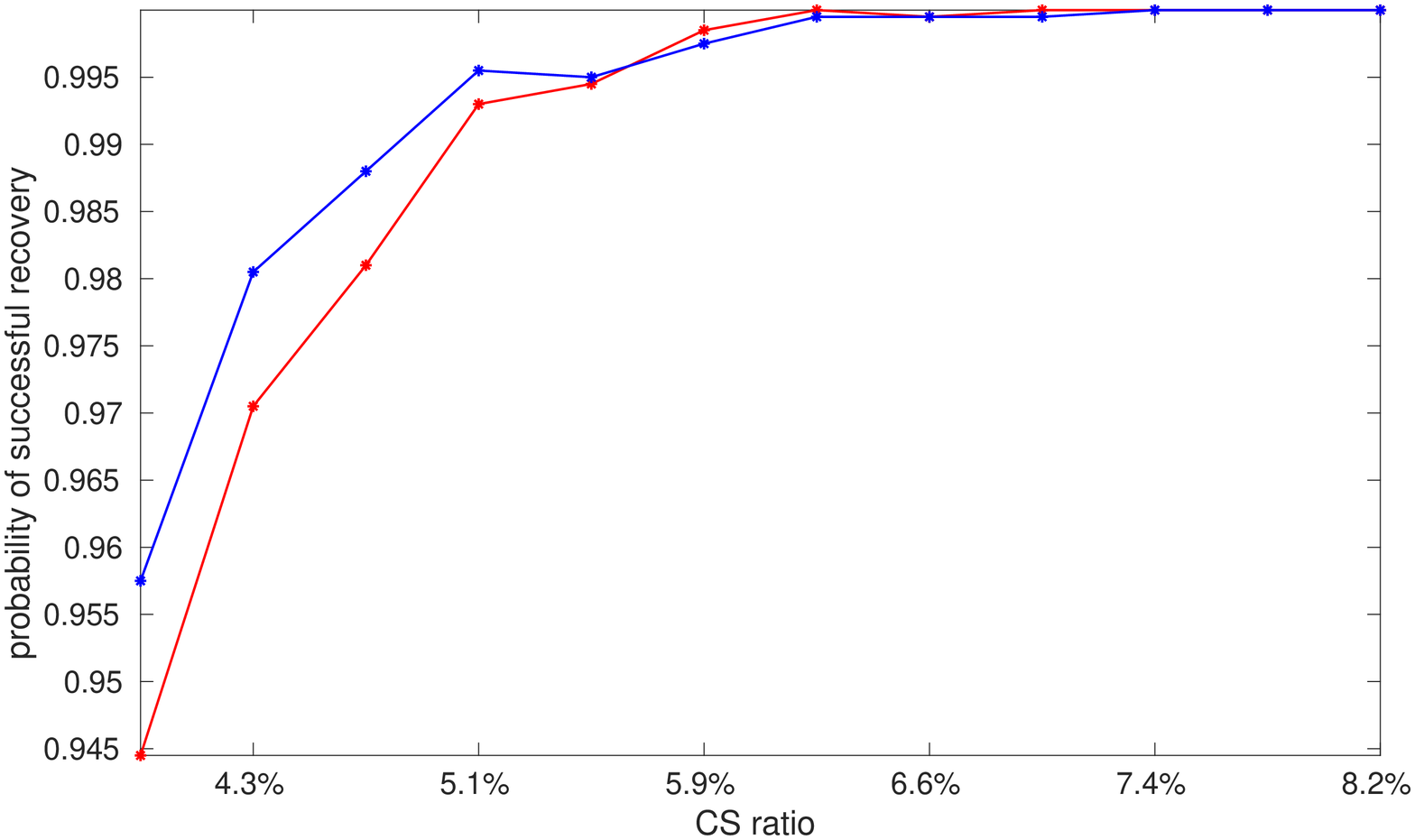,width=8cm}}}
\centerline{\hspace{0.8cm}(a1) k = 10 
\hspace{7cm}(a2) k = 10 }
\vspace{0.4cm}
{\centerline{\hspace{0cm}
\epsfig{figure=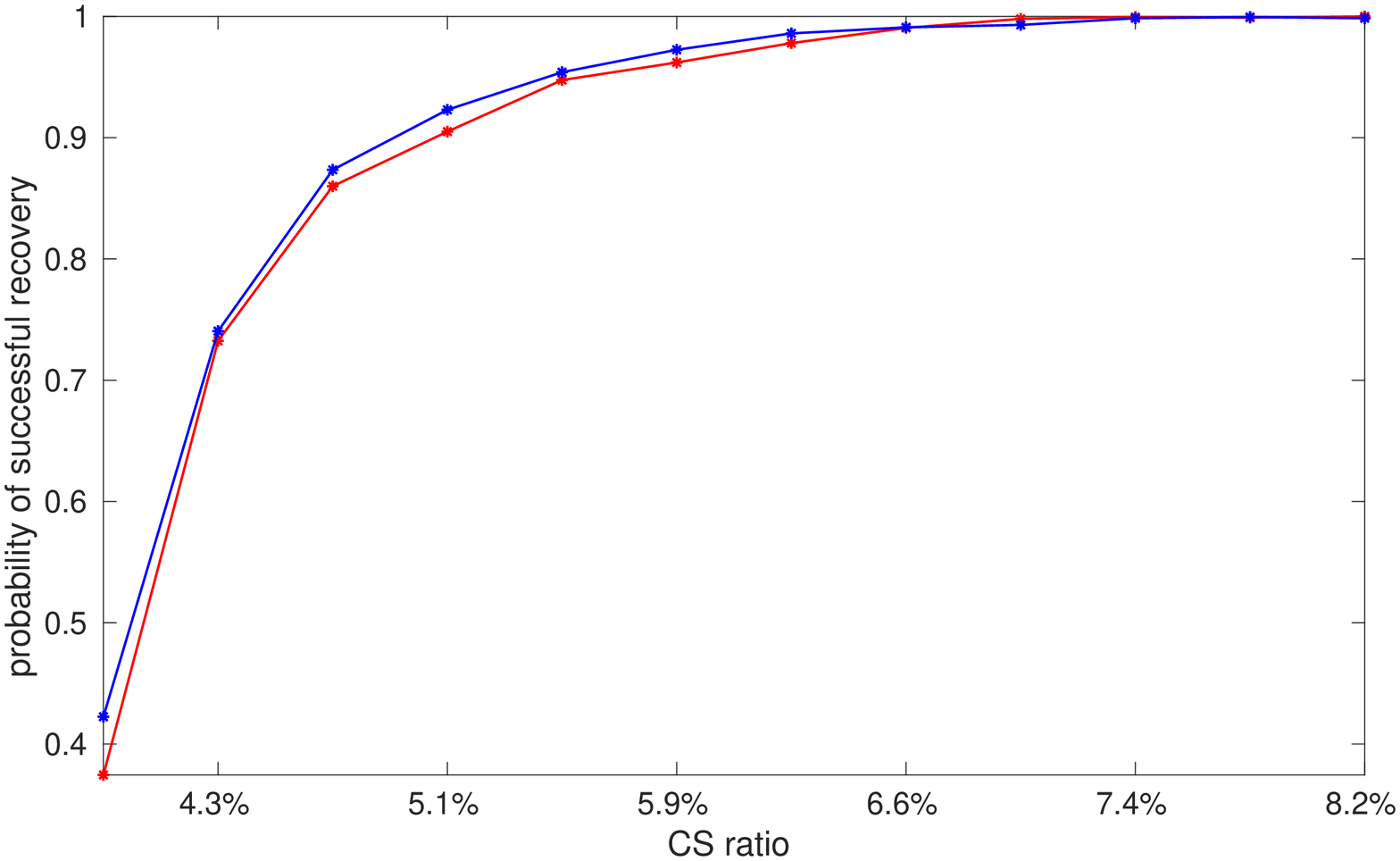,width=8cm}\hspace{0.4cm}
\epsfig{figure=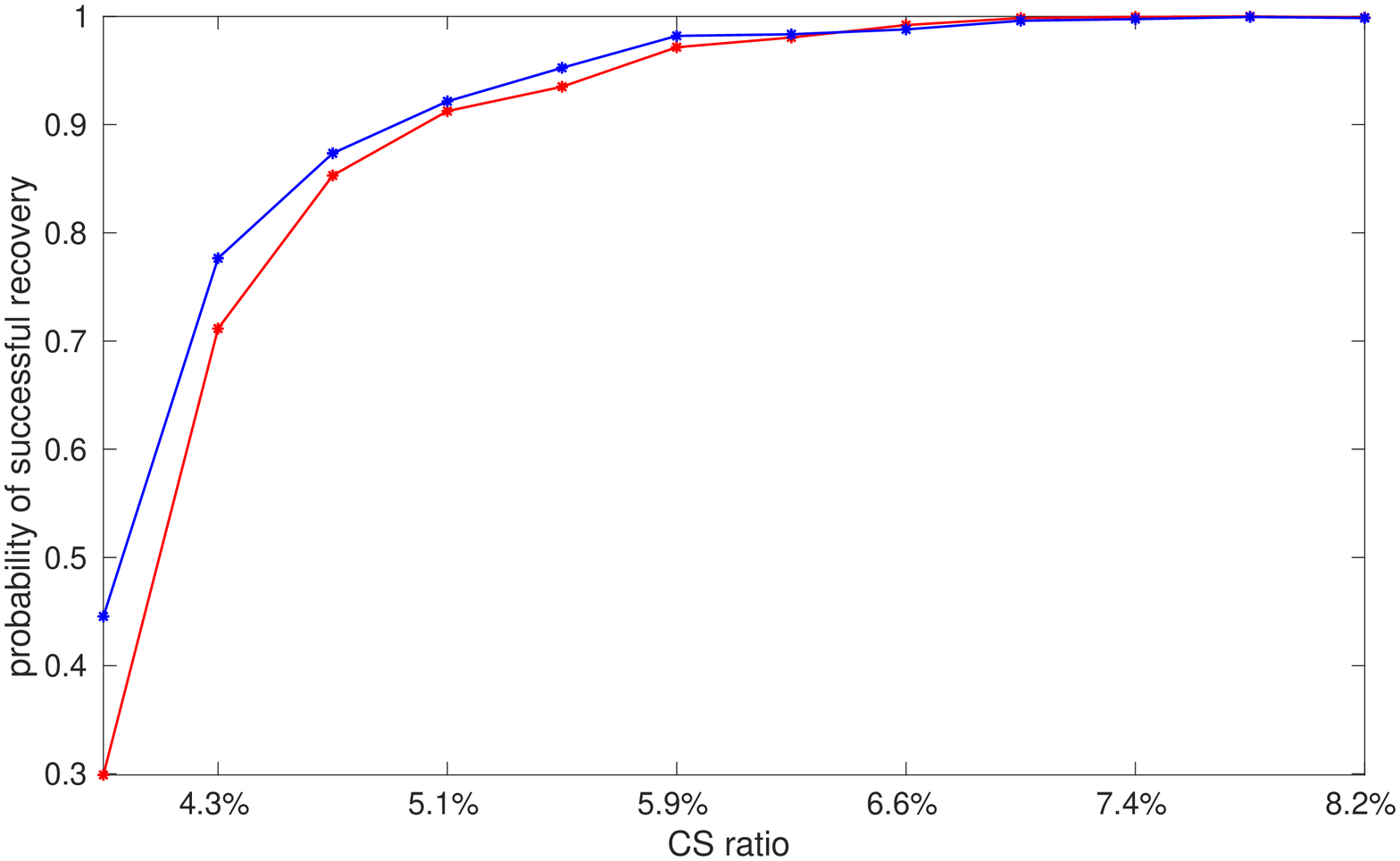,width=8cm}}}
\centerline{\hspace{0.8cm}(b1)  k = 12
\hspace{7cm}(b2) k = 12 }
\vspace{0.4cm}
{\centerline{\hspace{0cm}
\epsfig{figure=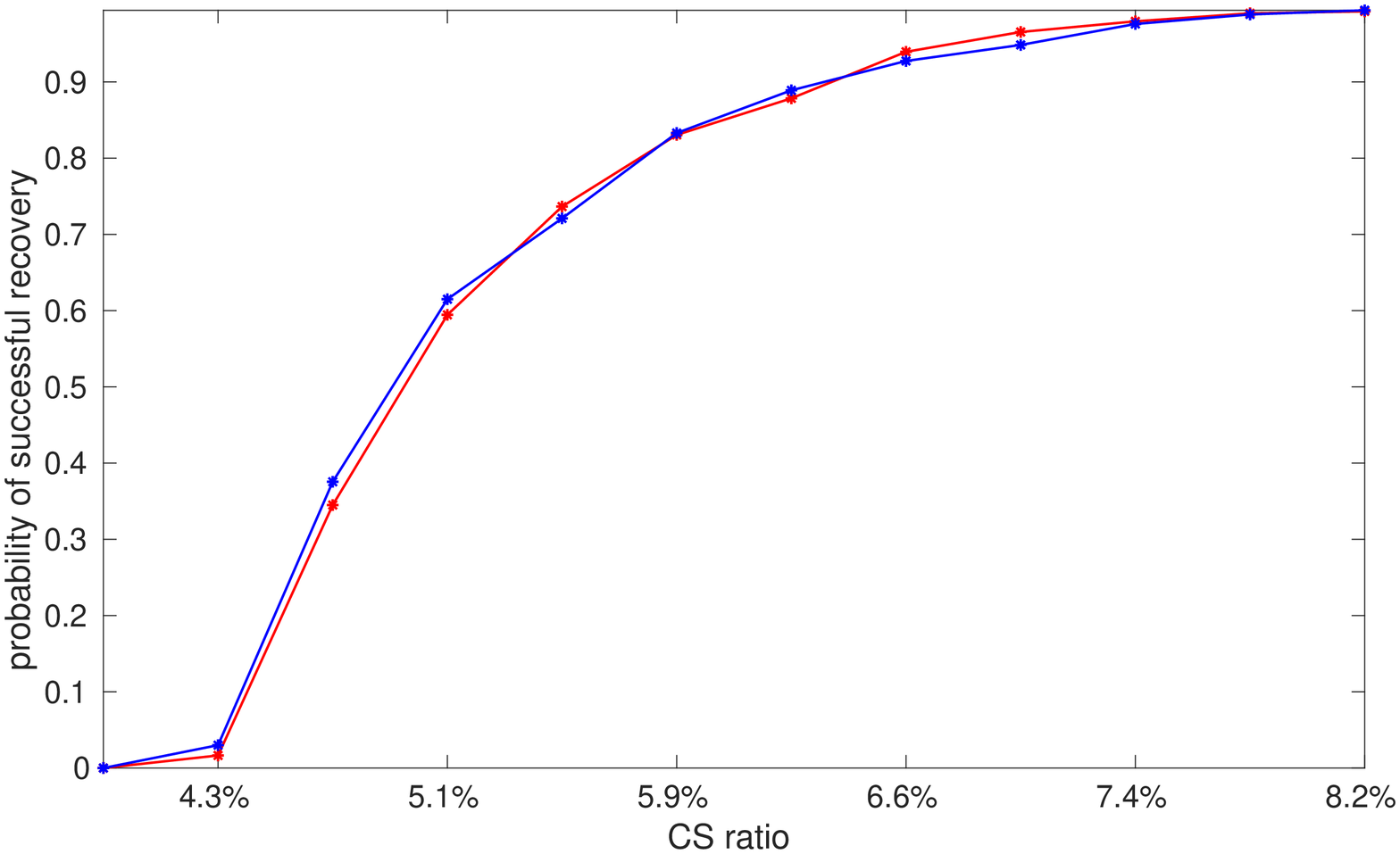,width=8cm}\hspace{0.4cm}
\epsfig{figure=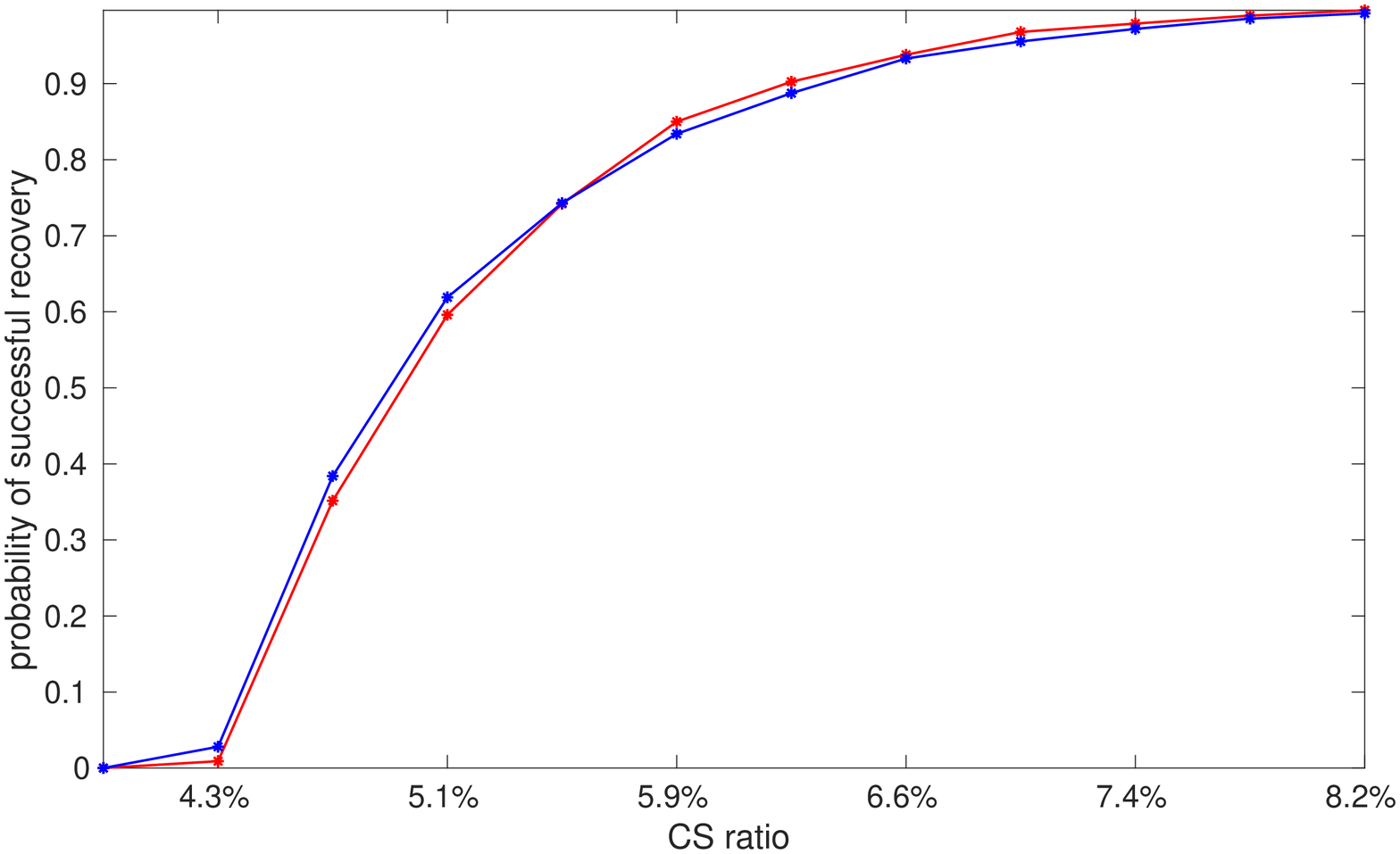,width=8cm}}}
\centerline{\hspace{0.8cm}(c1) k = 14
\hspace{7cm}(c2) k = 14 }
\caption{\label{fig_KSVD}
Comparisons of CS recovery performance (i.e., the probability of sparse vector recovery versus CS ratio) using  sparsifying dictionary $D_{KSVD}$. Red and blue curves were respectively obtained using the benchmark (\ref{cs_p2}) and our approach (\ref{cs_p1}). Sparse vectors $x_i$ were randomly generated and each point on the curve is the average of $2,000$ probability measurements. The positions of non-zero coefficients of $x_i$ are uniformly distributed and the values of the non-zero coefficients of $x_i$ are uniformly distributed in $[-1, 1]$.
In (a1), (b1), and (c1), $A$ is a Gaussian random matrix. 
In (a2), (b2), and (c2), $A$ is a Bernoulli random matrix. }
\end{figure}

\begin{figure}[h!]
{\centerline{\hspace{0cm}
\epsfig{figure=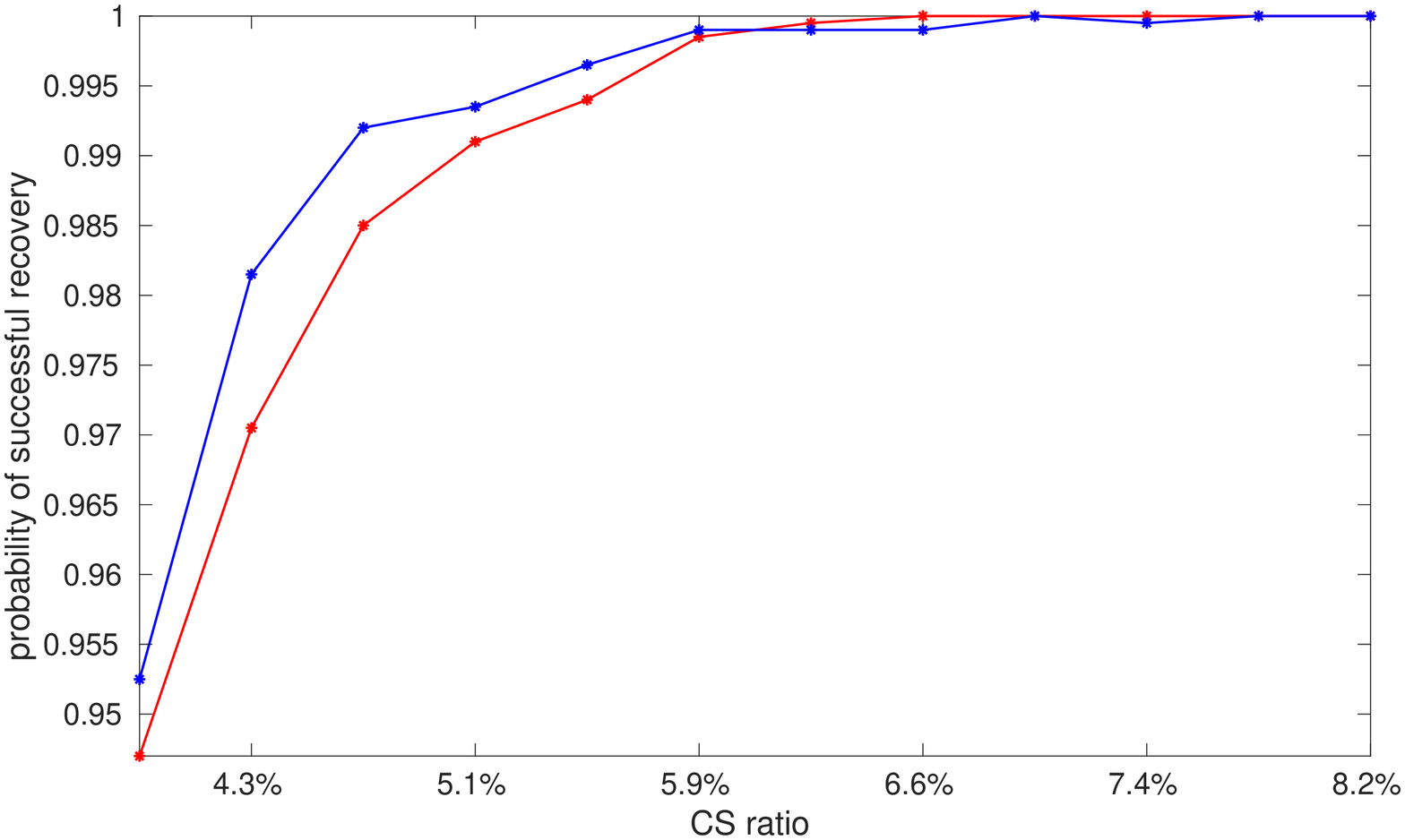,width=8cm}\hspace{0.4cm}
\epsfig{figure=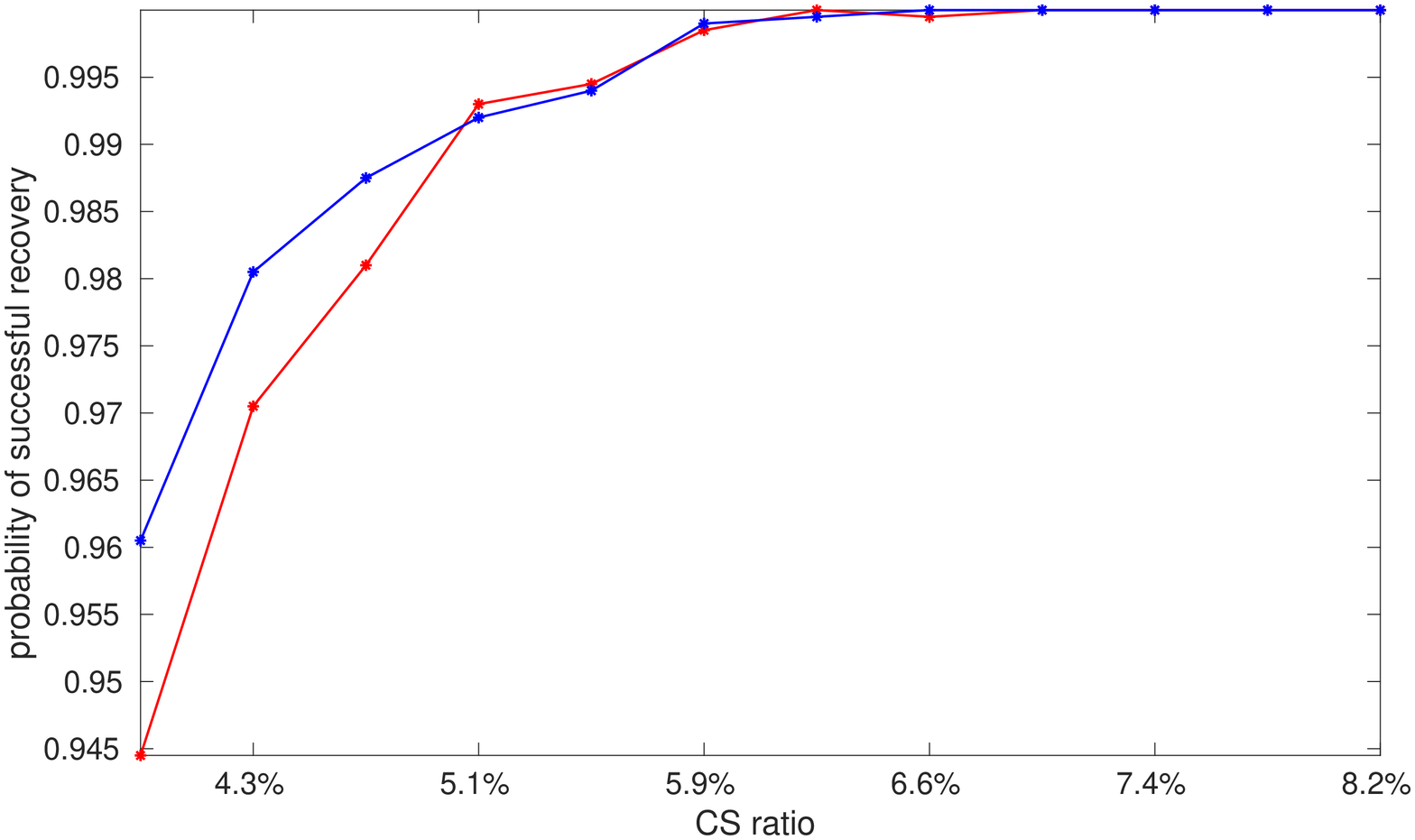,width=8cm}}}
\centerline{\hspace{0.8cm}(a1) k = 10
\hspace{7cm}(a2) k = 10}
\vspace{0.3cm}
{\centerline{\hspace{0cm}
\epsfig{figure=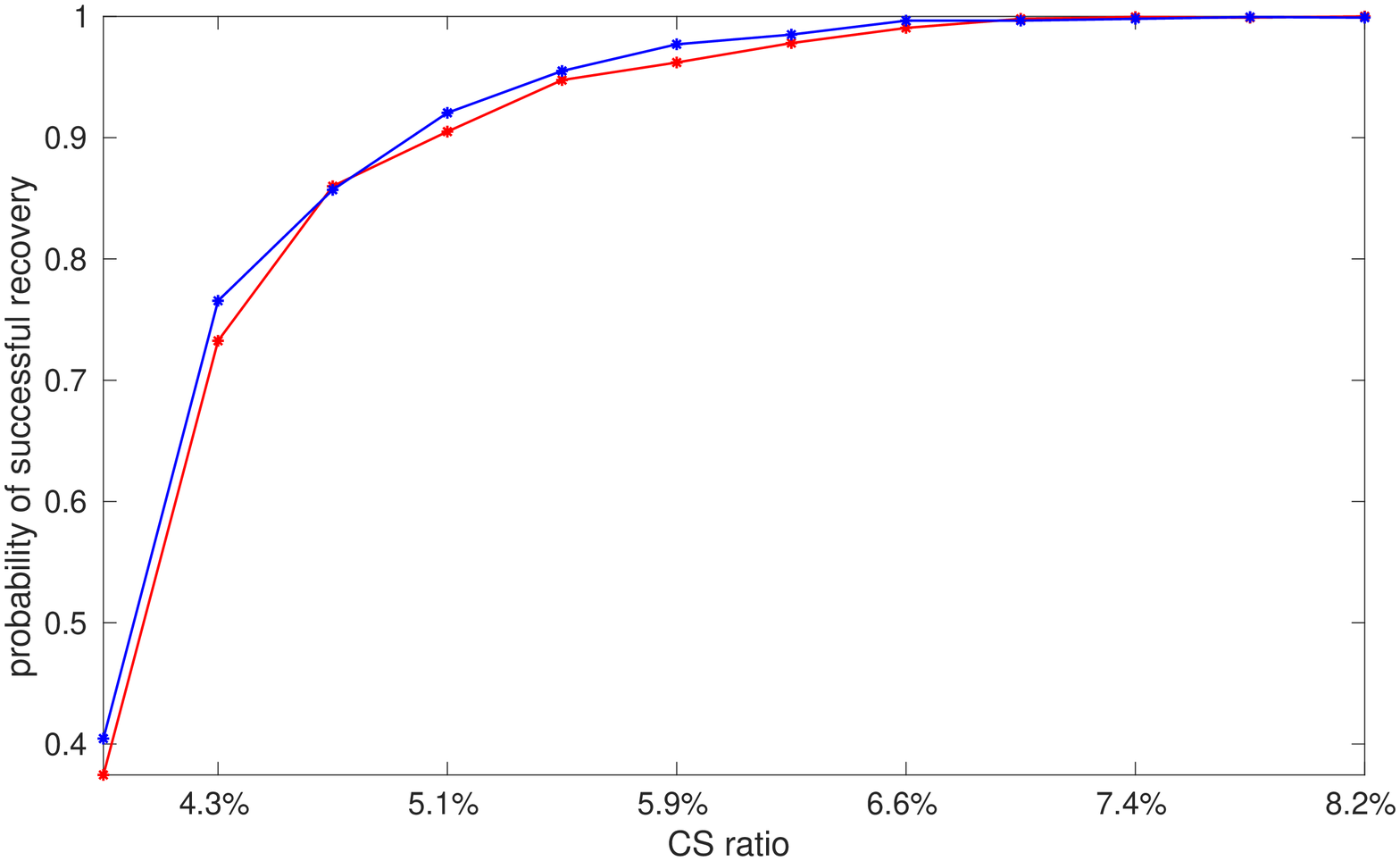,width=8cm}\hspace{0.4cm}
\epsfig{figure=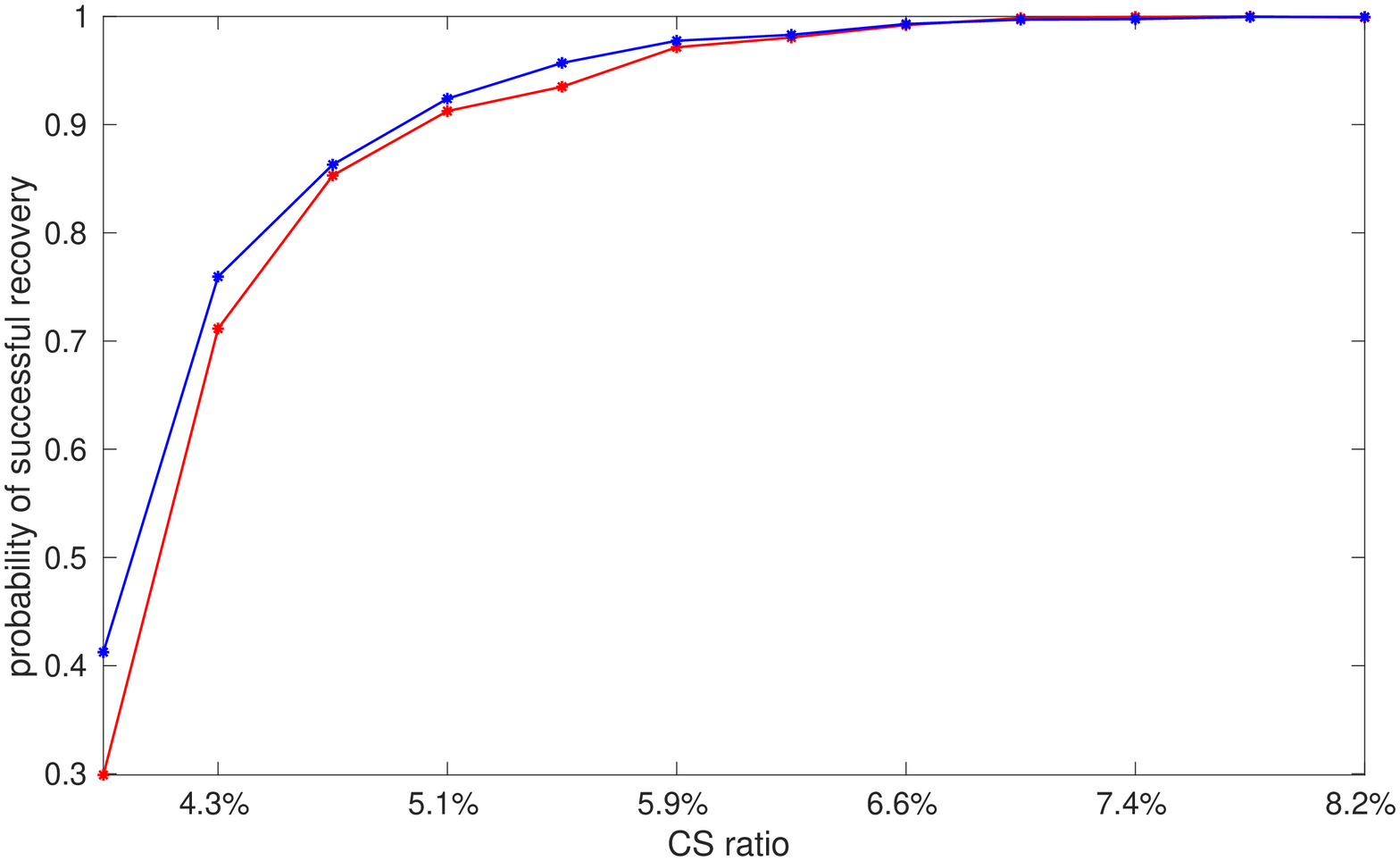,width=8cm}}}
\centerline{\hspace{0.8cm}(b1) k = 12
\hspace{7cm}(b2) k = 12}
\vspace{0.3cm}
{\centerline{\hspace{0cm}
\epsfig{figure=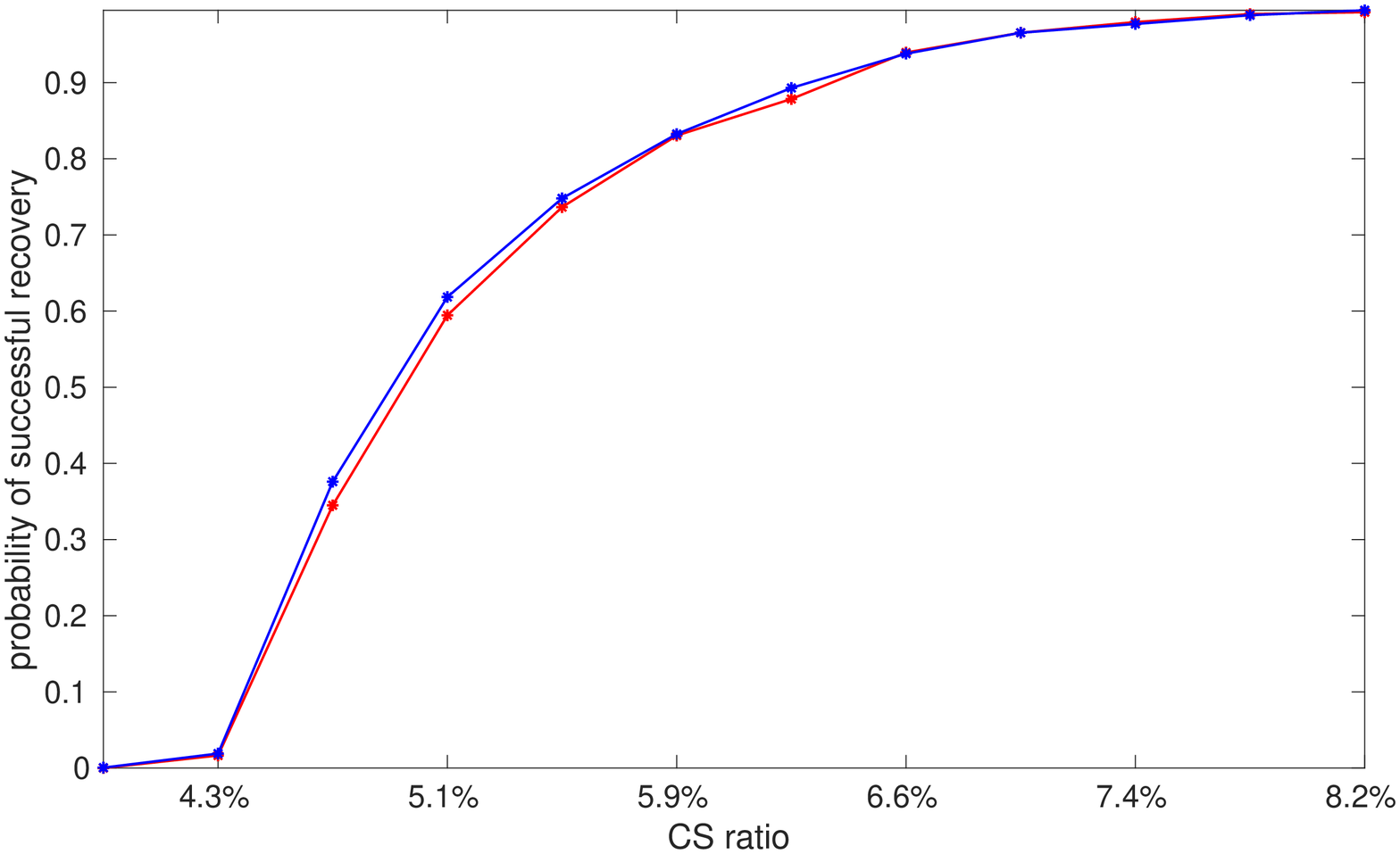,width=8cm}\hspace{0.4cm}
\epsfig{figure=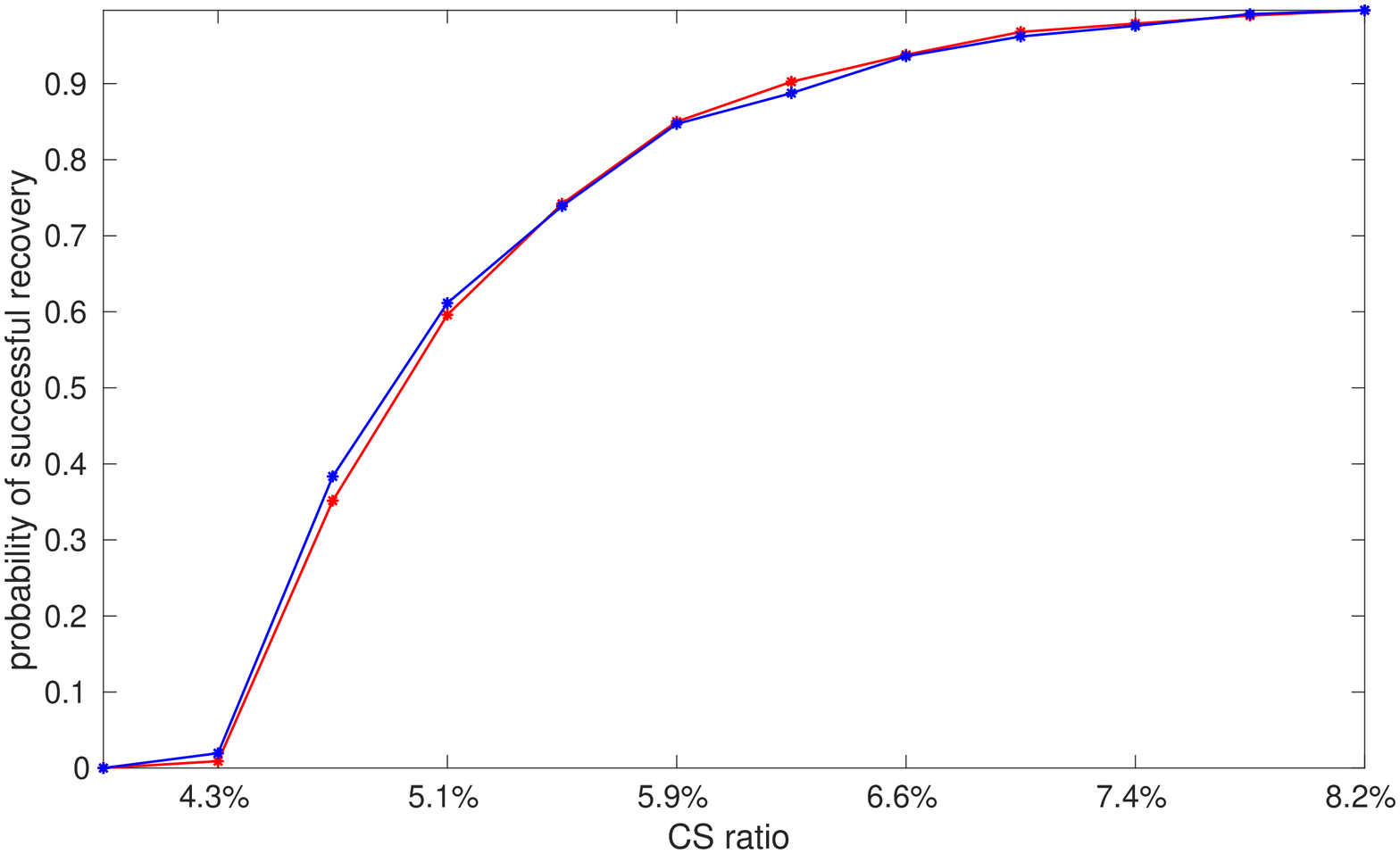,width=8cm}}}
\centerline{\hspace{0.8cm}(c1) k= 14
\hspace{7cm}(c2) k= 14}
\caption{\label{fig_PKSVD}
Comparisons of CS recovery performance (i.e., the probability of sparse vector recovery versus CS ratio) using  sparsifying dictionary $D_{PKSVD}$. Red and blue curves were respectively obtained using the benchmark (\ref{cs_p2}) and our approach (\ref{cs_p1}). Sparse vectors $x_i$ were randomly generated and each point on the curve is the average of $2,000$ probability measurements. The positions of non-zero coefficients of $x_i$ are uniformly distributed and the values of the non-zero coefficients of $x_i$ are uniformly distributed in $[-1, 1]$.
In (a1), (b1), and (c1), $A$ is a Gaussian random matrix. 
In (a2), (b2), and (c2), $A$ is a Bernoulli random matrix. }
\end{figure}

\begin{figure}[h!]
{\centerline{\hspace{0cm}
\epsfig{figure=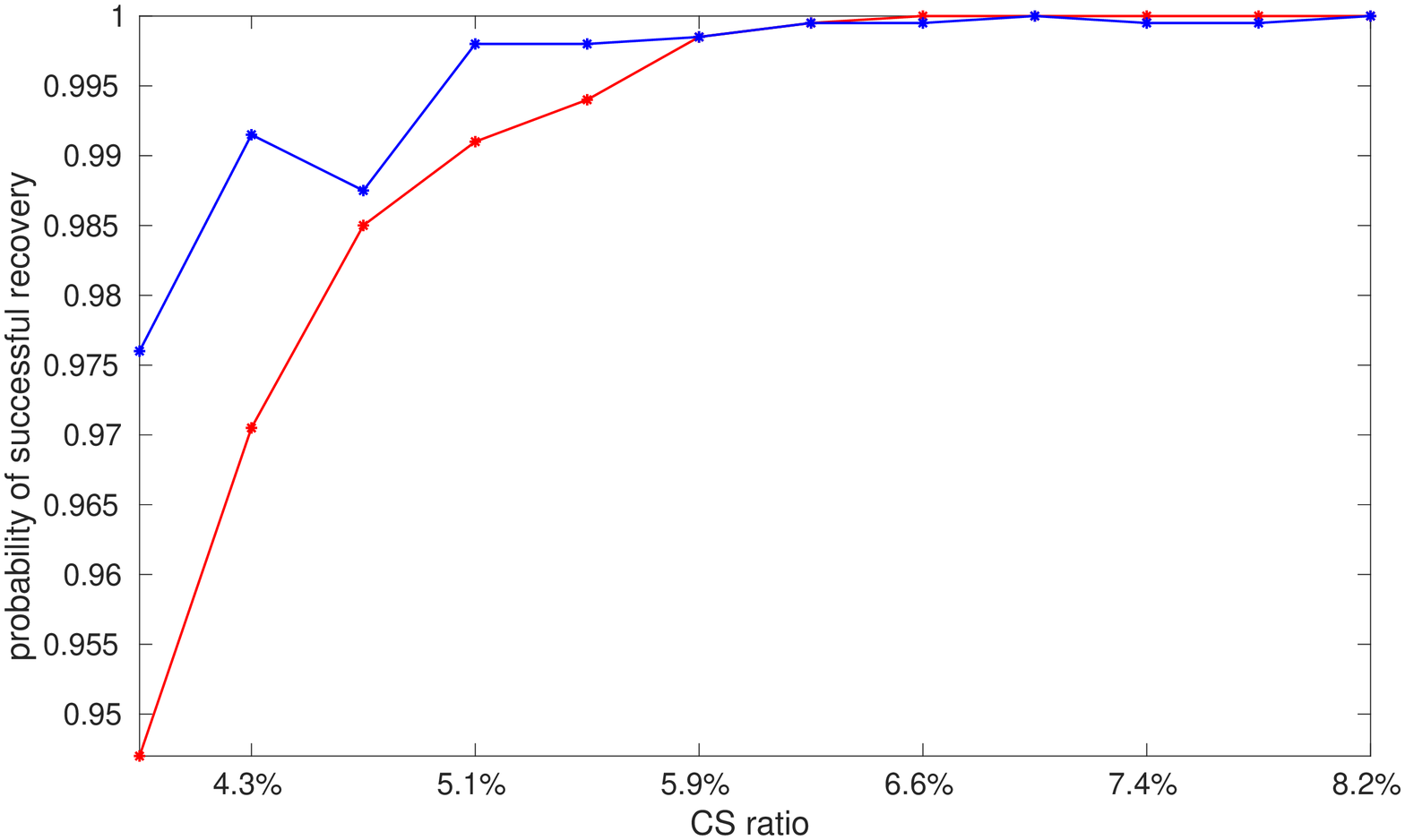,width=8cm}\hspace{0.4cm}
\epsfig{figure=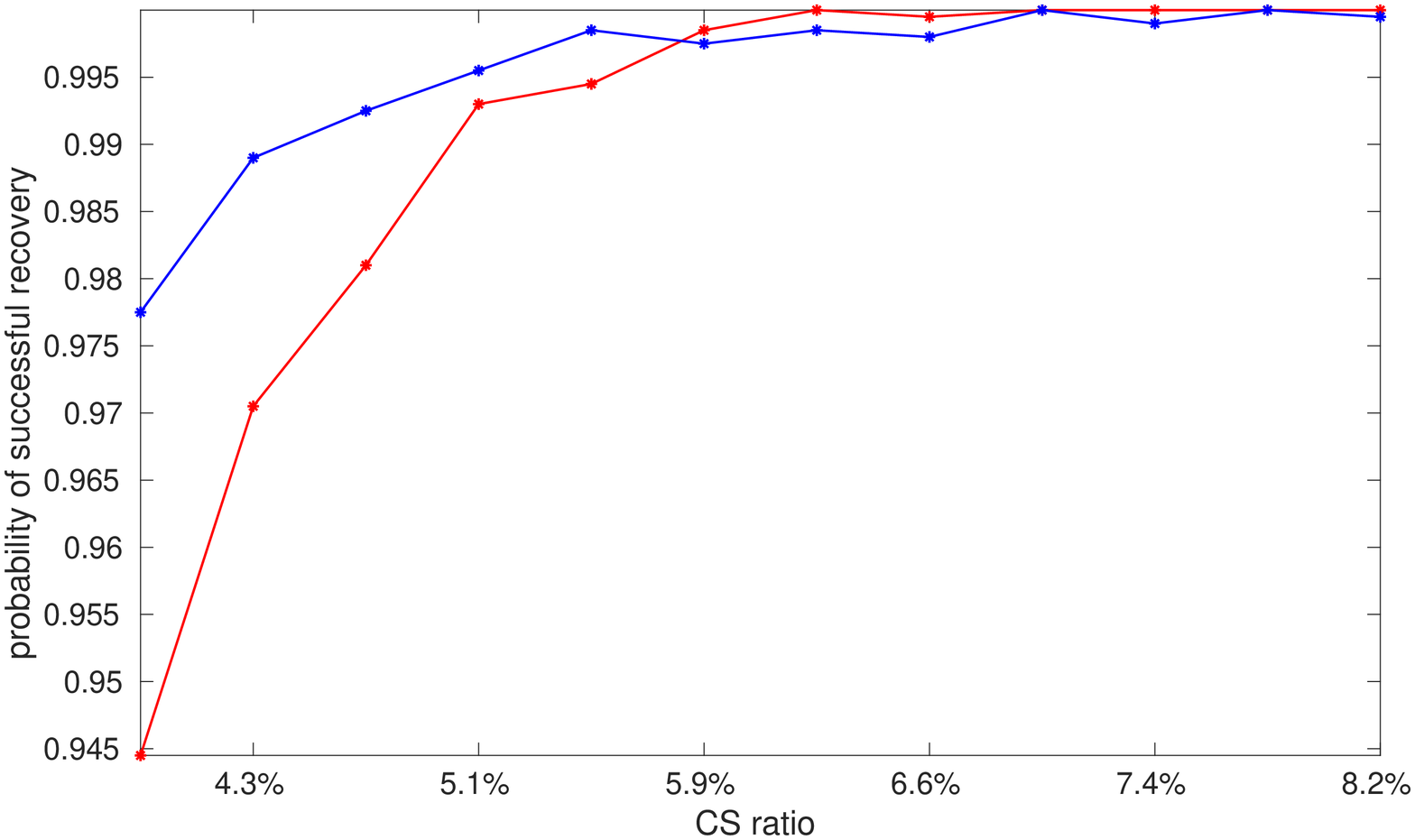,width=8cm}}}
\centerline{\hspace{0.8cm}(a1) k = 10 
\hspace{7cm}(a2) k = 10}
\vspace{0.3cm}
{\centerline{\hspace{0cm}
\epsfig{figure=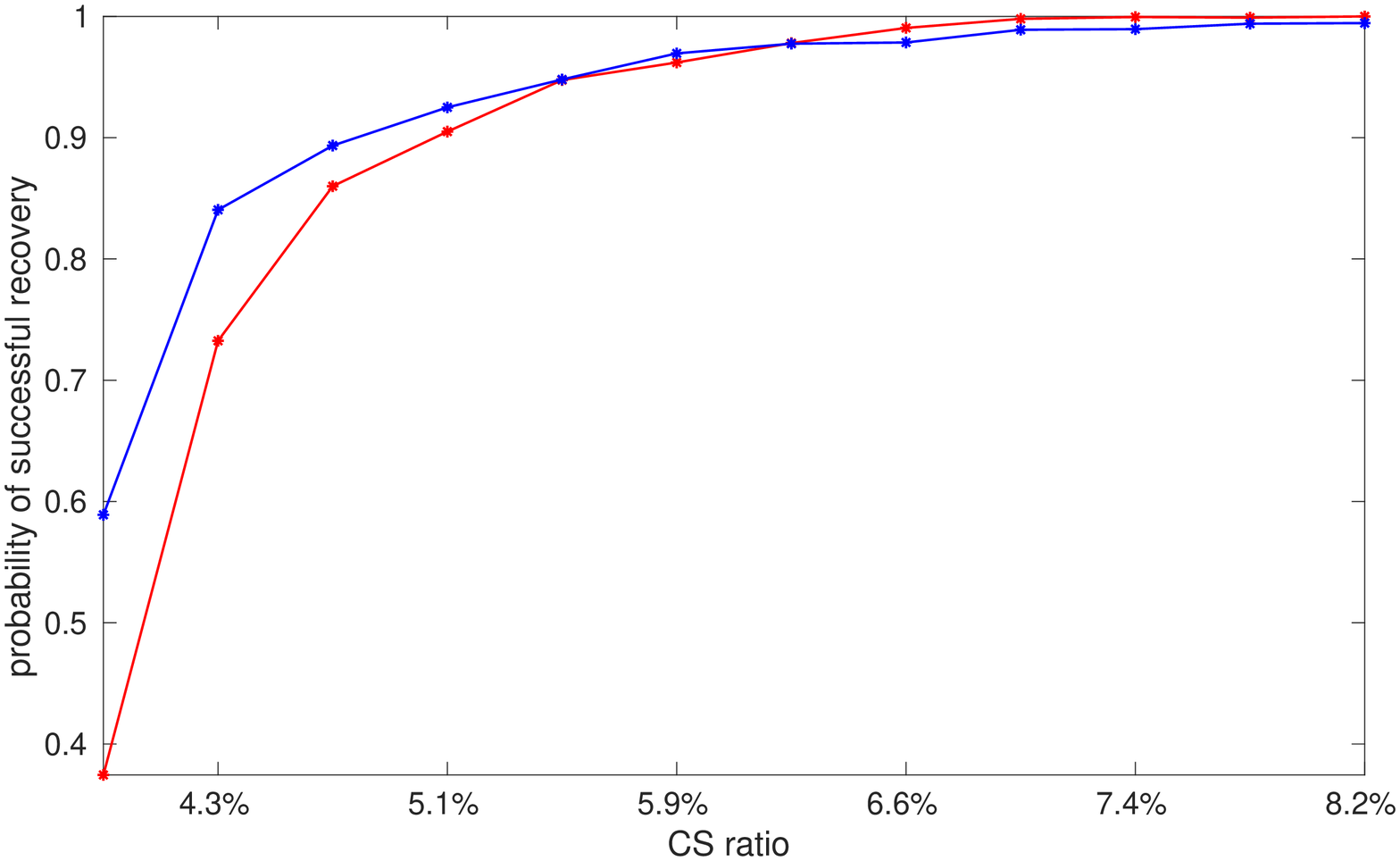,width=8cm}\hspace{0.4cm}
\epsfig{figure=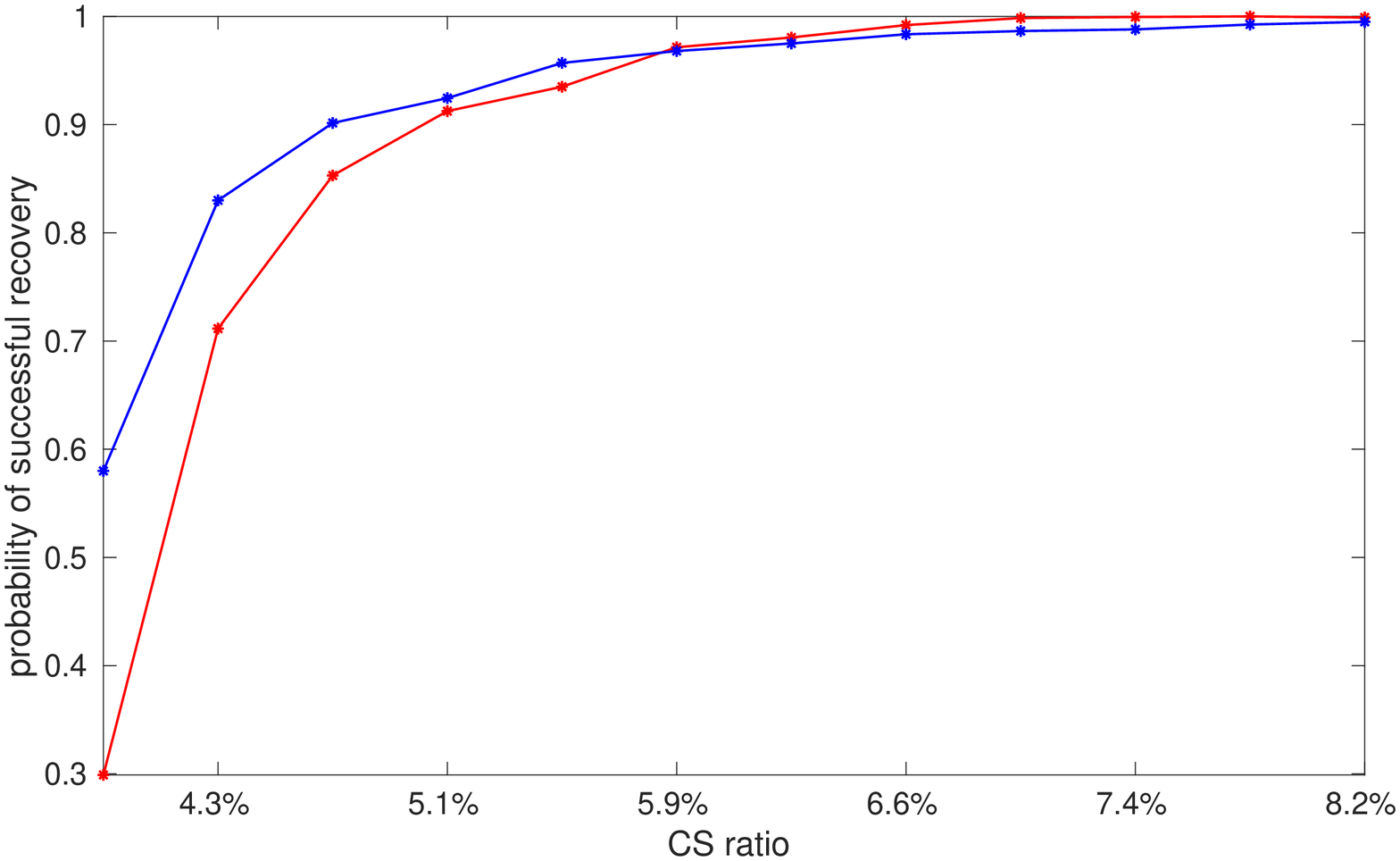,width=8cm}}}
\centerline{\hspace{0.8cm}(b1) k = 12
\hspace{7cm}(b2) k = 12}
\vspace{0.3cm}
{\centerline{\hspace{0cm}
\epsfig{figure=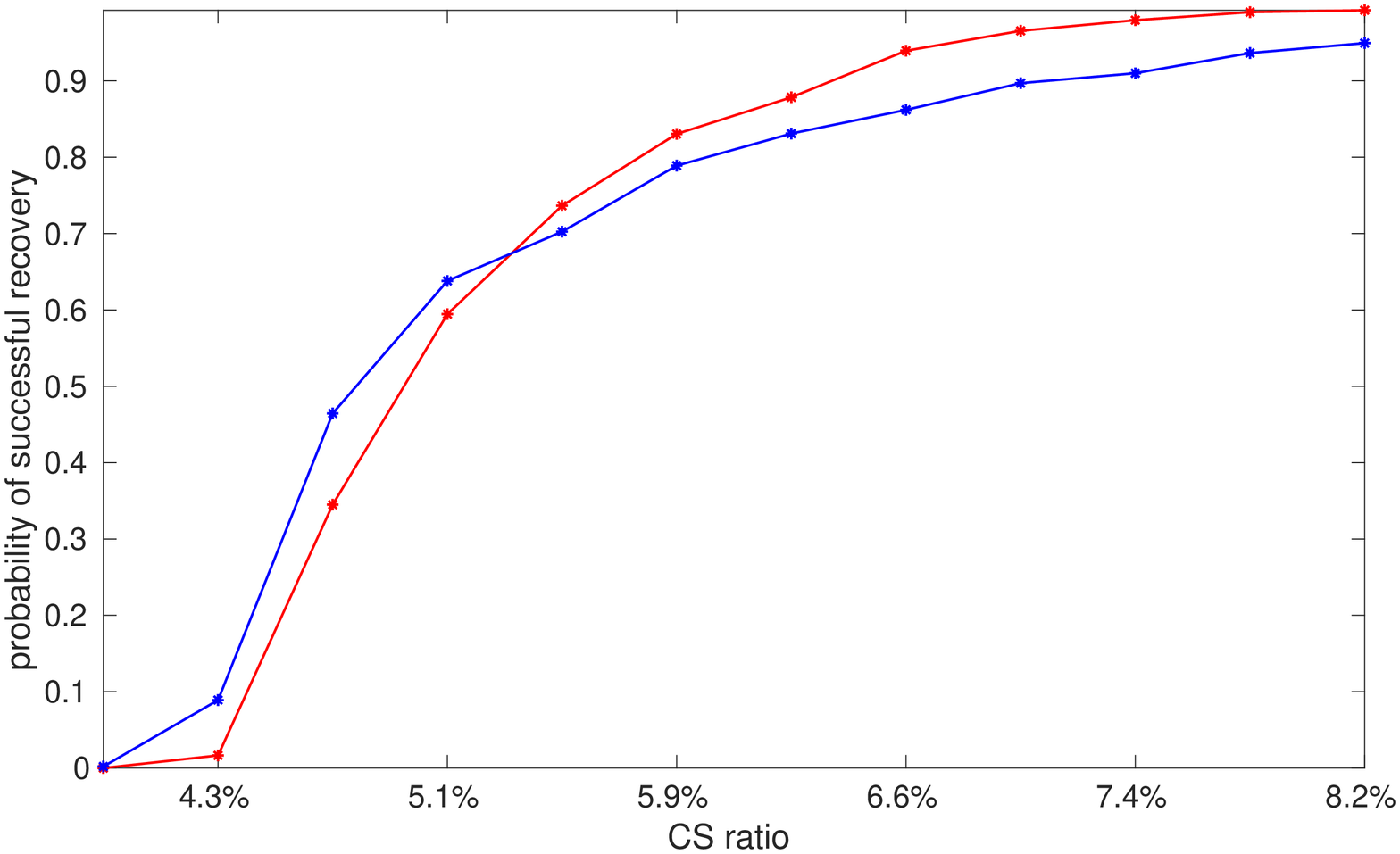,width=8cm}\hspace{0.4cm}
\epsfig{figure=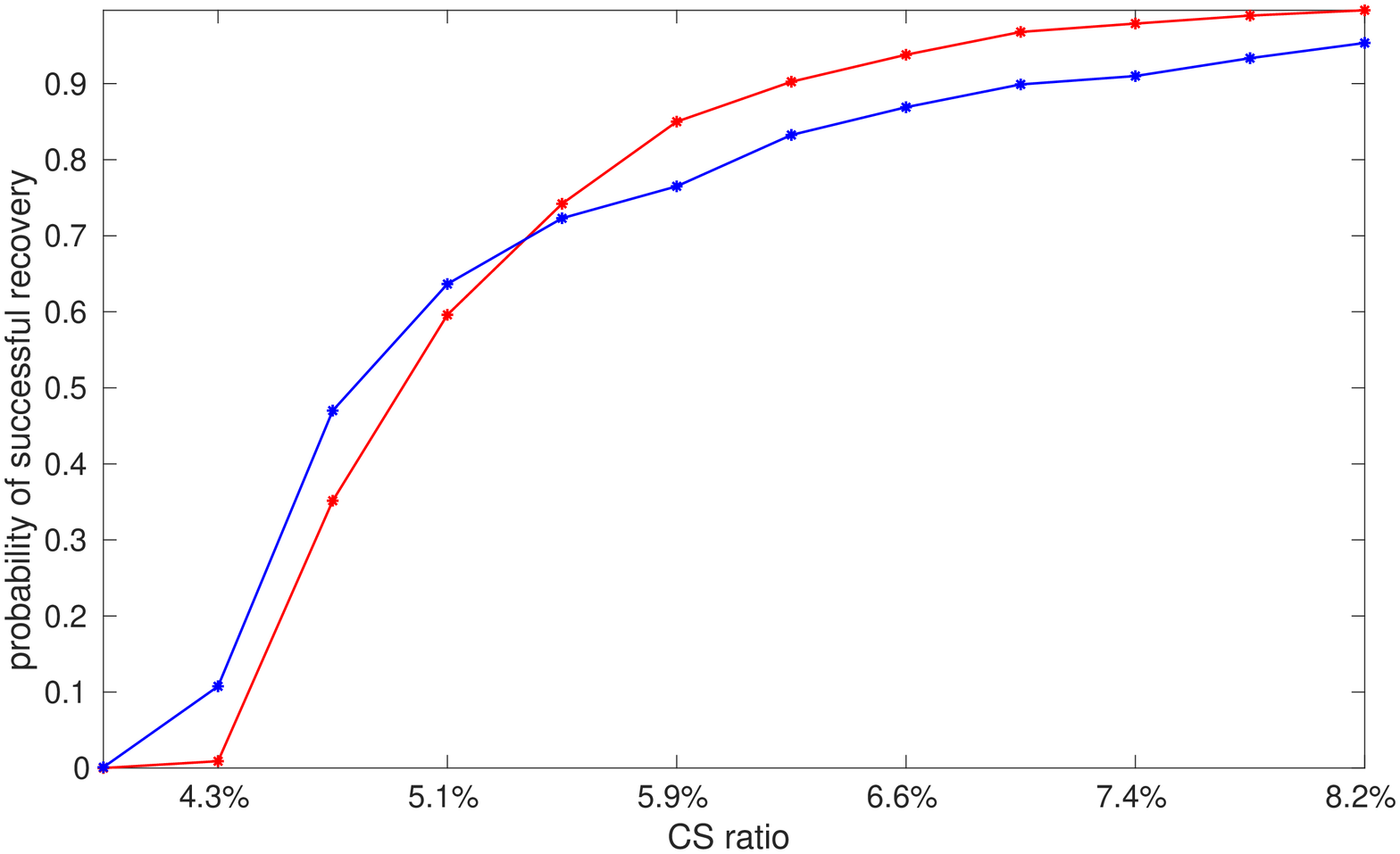,width=8cm}}}
\centerline{\hspace{0.8cm}(c1) k = 14
\hspace{7cm}(c2) k = 14}
\caption{\label{fig_wavelet}
Comparisons of CS recovery performance (i.e., the probability of sparse vector recovery versus CS ratio) using  sparsifying dictionary $D_{wavelet}$. Red and blue curves were respectively obtained using the benchmark (\ref{cs_p2}) and our approach (\ref{cs_p1}). Sparse vectors $x_i$ were randomly generated and each point on the curve is the average of $2,000$ probability measurements. The positions of non-zero coefficients of $x_i$ are uniformly distributed and the values of the non-zero coefficients of $x_i$ are uniformly distributed in $[-1, 1]$.
In (a1), (b1), and (c1), $A$ is a Gaussian random matrix. 
In (a2), (b2), and (c2), $A$ is a Bernoulli random matrix. }
\end{figure}

Figures \ref{fig_KSVD}, \ref{fig_PKSVD}, and \ref{fig_wavelet} illustrate CS recovery performance via plots indicating the probability of successfully recovering a sparse vector versus the CS ratio. The figures compare the performance of our approach (\ref{cs_p1}) against the benchmark (\ref{cs_p2}) using Gaussian and Bernoulli sensing matrices at various levels of sparsity. 
The horizontal and vertical axes, respectively, indicate the CS ratio (i.e.,  $\frac{m}{n}\times 100 \%$, where $m$ is the number of rows of $\mathcal E_i$ and $n=1024$) and the probability of successfully recovering a sparse vector.
We claim that the true sparse vector $x$ can be recovered as long as the estimate $\hat x$ satisfies $\| \hat x - x \|_1  < 1024 \times 10^{-2}$.

\section{Conclusions} \label{conclusion}
\begin{figure}[h!]
{\centerline{\hspace{0.5cm}
\epsfig{figure=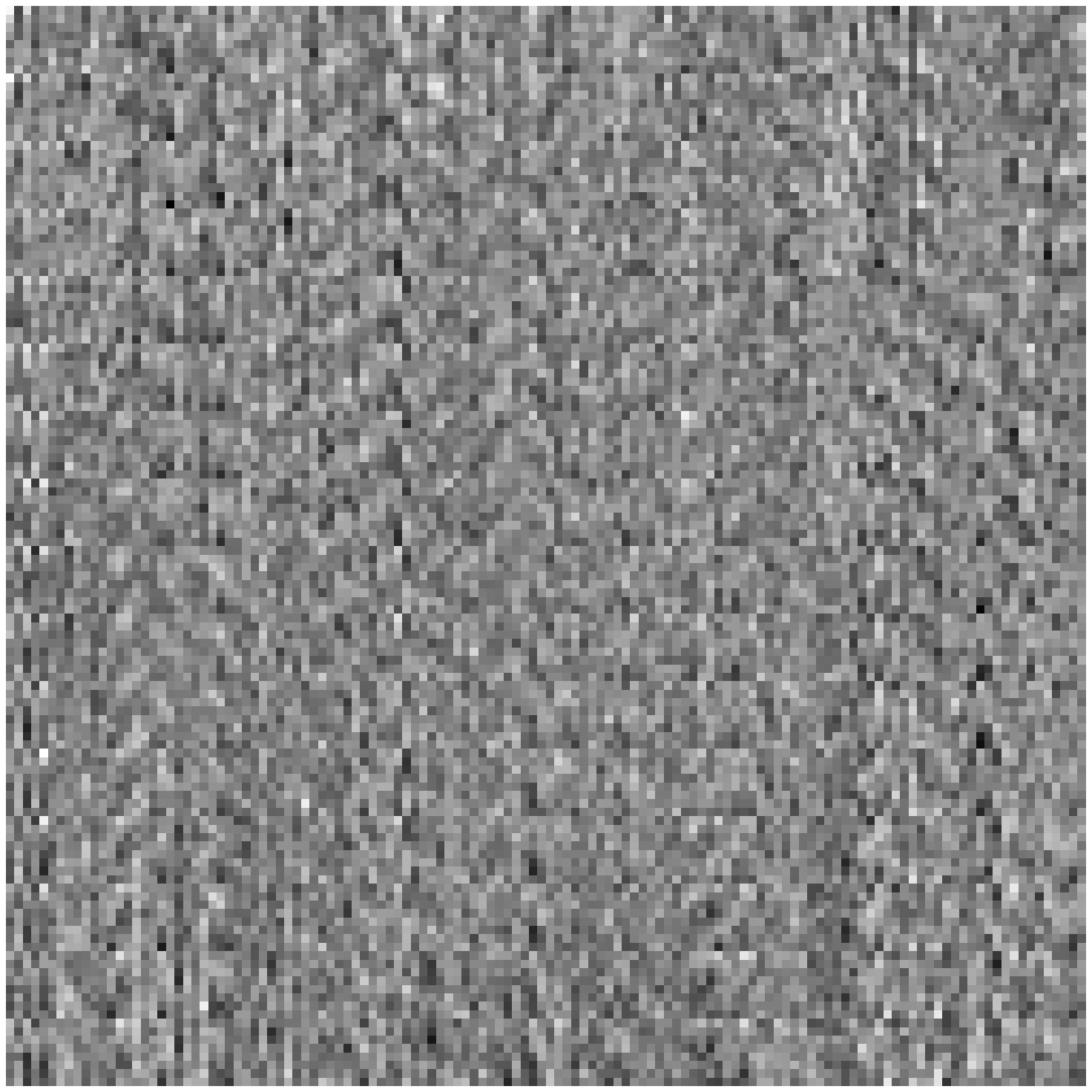,width=6cm}
\epsfig{figure=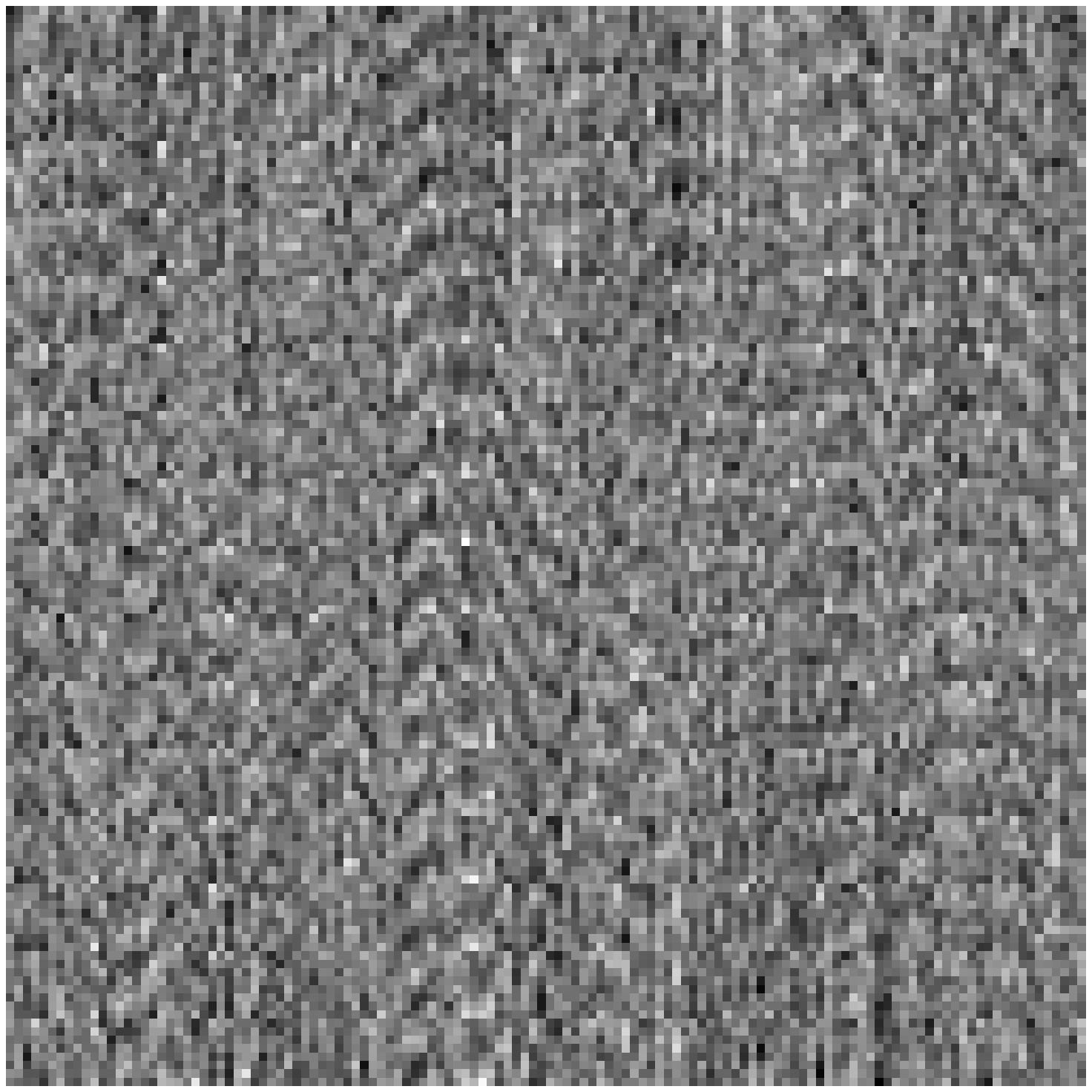,width=6cm}}}
\centerline{\hspace{0.8cm}(a1) $G_{KSVD}$ (Gaussian)  
\hspace{2cm}(a2) $G_{KSVD}$ (Bernoulli)}
{\centerline{\hspace{0.5cm}
\epsfig{figure=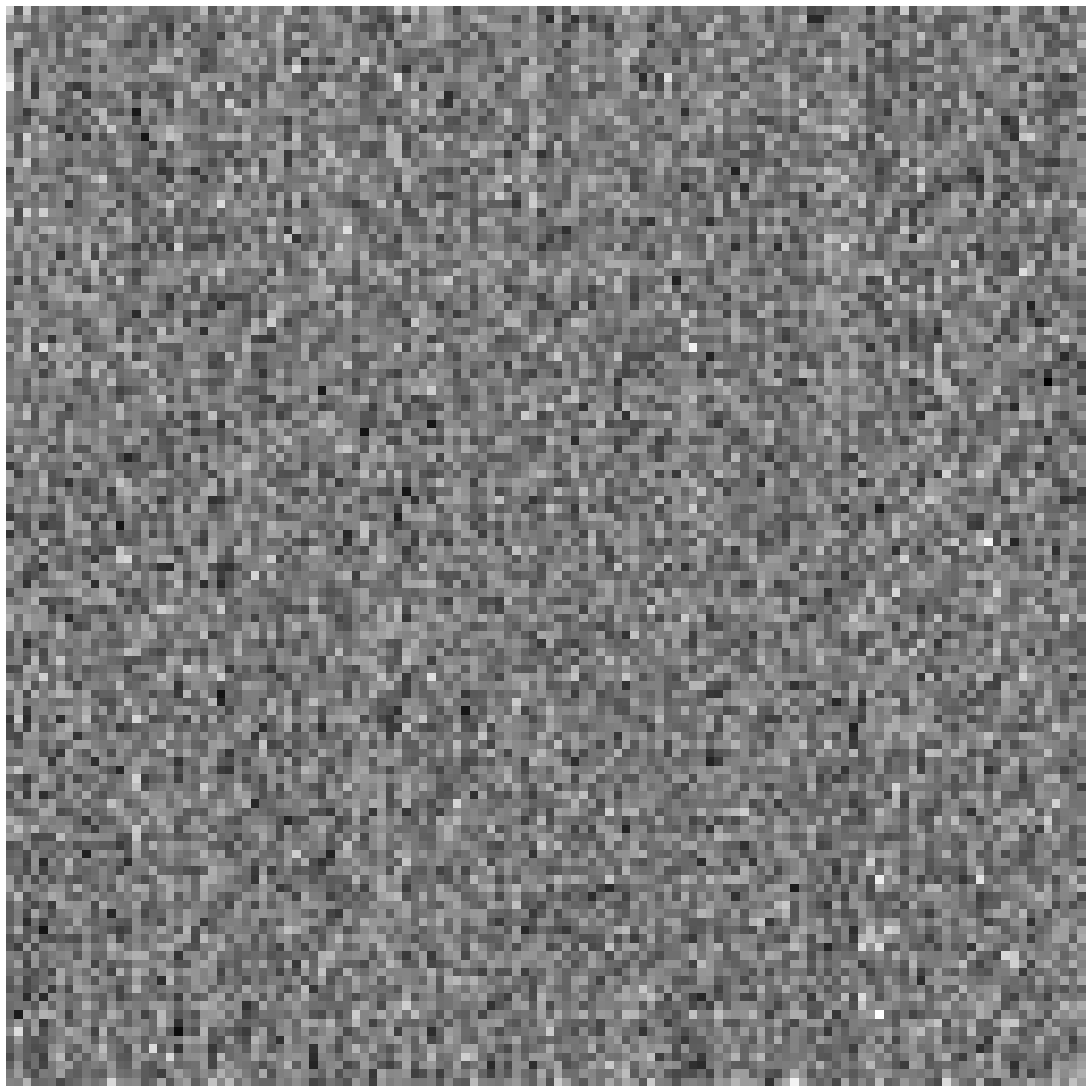,width=6cm}
\epsfig{figure=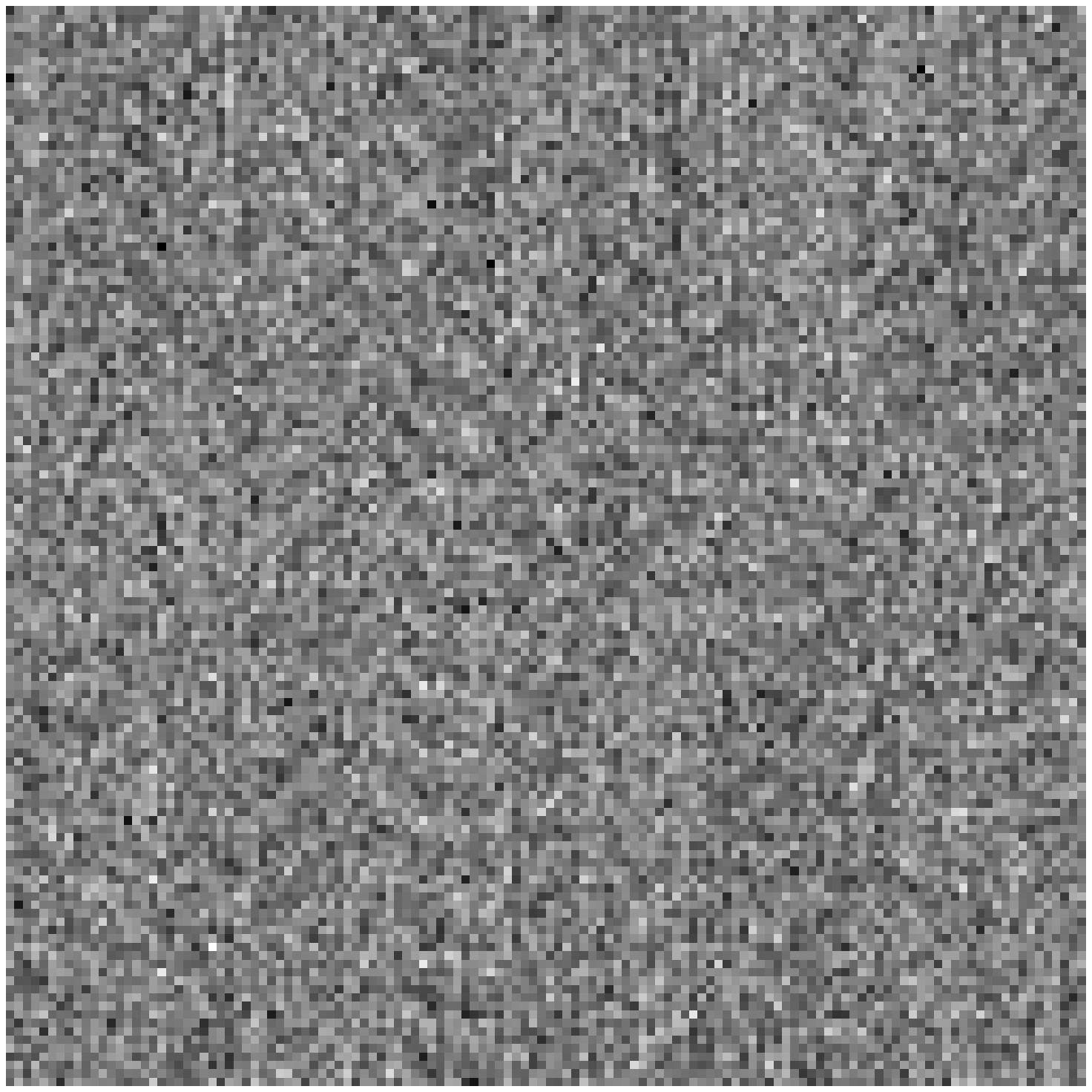,width=6cm}}}
\centerline{\hspace{0.8cm}(a1) $G_{PKSVD}$ (Gaussian)  
\hspace{2cm}(a2) $G_{PKSVD}$ (Bernoulli)}
{\centerline{\hspace{0.5cm}
\epsfig{figure=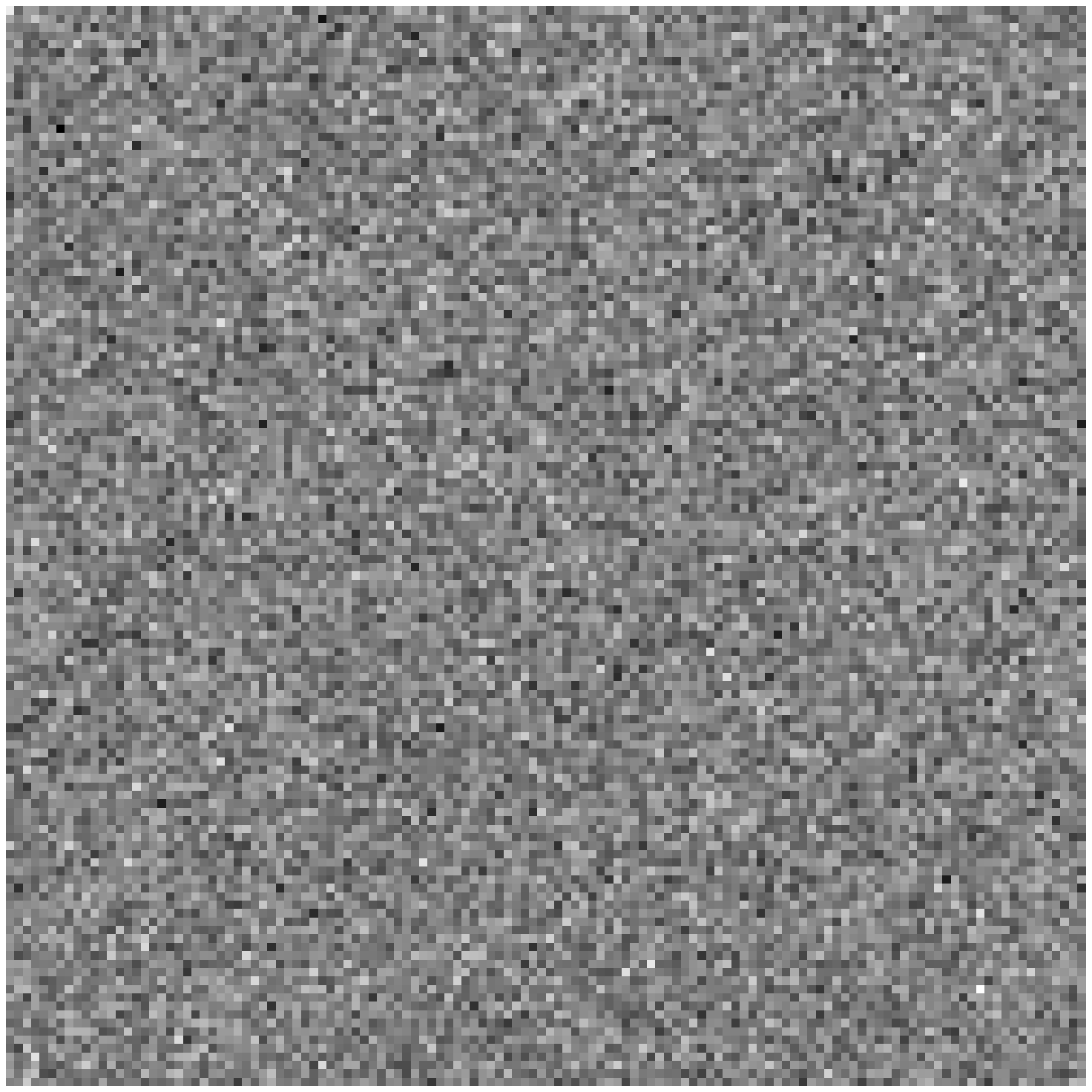,width=6cm}
\epsfig{figure=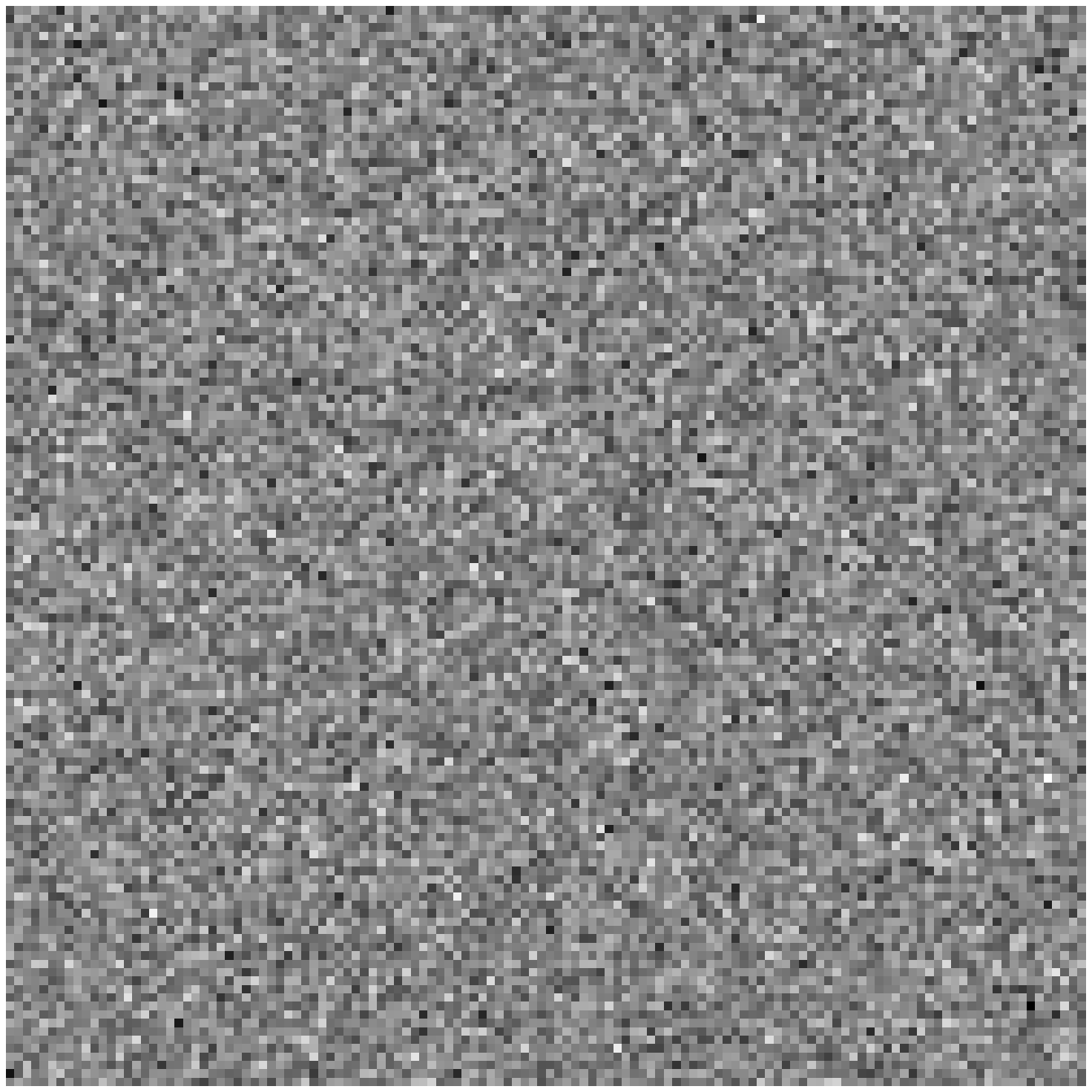,width=6cm}}}
\centerline{\hspace{0.8cm}(a1) $G_{wavelet}$ (Gaussian)  
\hspace{2cm}(a2) $G_{wavelet}$ (Bernoulli)}
\caption{\label{fig_G}
Matrix $G\in\mathbb{R}^{128\times128}$ of several sparsifying dictionaries $D$ for various $A$. Recall that $D = GAH$ and the sensing matrix of $D$ is $\mathcal E G^{-1}$. 
In (a1), (b1), and (c1), $A$ is a Gaussian random matrix. 
In (a2), (b2), and (c2), $A$ is a Bernoulli random matrix. 
In (a1) and (a2), $D = D_{KSVD}$.
In (b1) and (b2), $D = D_{PKSVD}$.
In (c1) and (c2), $D = D_{wavelet}$.
Note that the structure in the sub-figures is barely discernable.
}
\end{figure}

This paper outlines a novel approach to CS involving the construction of sensing matrix for a sparsifying dictionary, such that the composition of sensing matrix and dictionary has the RIP. The proposed approach is based on  factorizing a sparsifying dictionary $D$ with the help of a matrix $A$ of equal rank, and satisfying the concentration inequality to achieve $D=GAH$. The factorization solutions for $G$ and $H$ are not unique, and $G$ is related to the sensing matrix of $D$.
The non-uniqueness raises the question which factorization solution would provide the most stable result, as indicated by the ability to recover the sparse vector by solving the inverse problem of $AH$ via convex optimization. As shown in Figure \ref{fig_G}, a lack of structure in $G$ raises concerns pertaining to the applicability of this approach to large scale systems with restricted computation and storage resources.\\

\noindent{\bf{Acknowledgements: }} The original proof of Lemma \ref{PROPOSITION_1} and Theorem \ref{tightframe} were shortened by Assist Prof Andreas Heinecke, National University of Singapore. We truly appreciate his comments and suggestions on the presentation of this paper.

\bibliographystyle{ieeetr}
\bibliography{piggybacking}

\end{document}